\documentclass[a4paper,12pt]{article}
\usepackage[utf8]{inputenc}
\usepackage{amsmath}
\usepackage{bm,upgreek}
\usepackage{graphicx}
\usepackage[font={small}, labelfont={bf,small}]{caption}

\DeclareCaptionLabelFormat{sfiglab}{\textbf{Figure S#2}}
\DeclareCaptionLabelFormat{stablab}{\textbf{Table S#2}}

\usepackage{booktabs}
\newcommand{\minitab}[2][l]{\begin{tabular}{#1}#2\end{tabular}}
\usepackage{multirow}

\usepackage{xr-hyper}

\usepackage{hyperref}
\hypersetup{hidelinks}

\usepackage{multirow}
\usepackage{rotating,tabularx}

\usepackage{scicite}
\usepackage{times}
\usepackage{amsfonts}
\usepackage[normalem]{ulem}
\usepackage{cancel}
\usepackage{xcolor}

\let\URL\url
\makeatletter
\def\url{\begingroup \catcode`\%=12\catcode`\#=12\relax\printurl}
\def\printurl#1{\@URL#1 \@nil\endgroup}
\def\@URL#1 #2\@nil{\URL{#1}\ifx\relax#2\relax \else; \url{#2\relax}\fi}
\makeatother

\usepackage{letltxmacro}
\makeatletter
\AtBeginDocument{%
  \@ifdefinable{\myorg@nameref}{%
    \LetLtxMacro\myorg@nameref\nameref
    \DeclareRobustCommand*{\nameref}[1]{%
      \emph{\myorg@nameref{#1}}%
    }%
  }%
}
\makeatother

\newcommand{\bibinfo}[1]{}
\newcommand{\eprint}[1]{}

\newcommand{\matr}[1]{\mathbf{#1}}     
\newcommand{\vect}[1]{\mathbf{#1}}     
\newcommand{\vectg}[1]{\bm{#1}} 

\newcommand{\Tmat}{\matr{T}}
\newcommand{\Smat}{\matr{S}}
\newcommand{\Pmat}{\matr{P}}
\newcommand{\pvec}{\vect{p}}

\newcommand{\rev}[1]{#1}
\newcommand{\revv}[1]{#1}

\newenvironment{sciabstract}{%
\begin{quote} \bf}
{\end{quote}}

\title{Flow stability for dynamic community detection}

\author{Alexandre Bovet$^{1,2,\ast}$, Jean-Charles Delvenne$^\text{2,3}$, Renaud Lambiotte$^\text{1}$\\
\normalsize{$^{\text{1}}$Mathematical Institute, University of Oxford, UK}\\
\normalsize{$^{\text{2}}$ICTEAM,  Université  catholique  de  Louvain,  Belgium}\\
\normalsize{$^{\text{3}}$CORE,  Université   catholique  de  Louvain,  Belgium}\\
\normalsize{$^{\ast}$To whom correspondence should be addressed;}\\ \normalsize{E-mail: \texttt{alexandre.bovet@maths.ox.ac.uk}}
}

\date{}

\begin{document}


\maketitle

\begin{sciabstract}

Many systems exhibit complex temporal dynamics due to the presence of different processes taking place simultaneously. An important task in such systems is to extract a simplified view of their time-dependent network of interactions. Community detection in temporal networks usually relies on aggregation over time windows or consider sequences of different stationary epochs. For dynamics-based methods, attempts to generalize static-network methodologies also face the fundamental difficulty that a stationary state of the dynamics does not always exist. Here, we derive a method based on a dynamical process evolving on the temporal network. Our method allows dynamics that do not reach a steady state and uncovers two sets of communities for a given time interval that accounts for the ordering of edges in forward and backward time. We show that our method provides a natural way to disentangle the different dynamical scales present in a system with synthetic and real-world examples.
\end{sciabstract}

\section*{Introduction}

Interactions in complex systems typically result from a multitude 
of temporal processes such as adaptation, cascading behaviour, or cyclical patterns
that all take place simultaneously but often at different spatial and temporal scales\cite{Bar-Yam2019}.
The concept of temporal 
networks\cite{Holme2013,Masuda2016,Holme2019,Porter2020} is used to study such 
time-dependent networks.
The fundamental constituent of temporal networks are events, instead 
of edges in the case \rev{of static networks}, that represent interactions between two nodes \rev{of a graph},
delimited in time, and usually take the form of a quadruplet  
\rev{$(u, v, s_i, e_i)$}, where $u$ is the source node, $v$ is the 
target node, \rev{$s_i$} is the starting time of event $i$ and \rev{\rev{$e_i$} is its ending time.}
\rev{Nodes of a network may represent, for example, individuals, companies, neurons, genes or words while events represent their relations which may refer to social interactions, economic transactions, activity correlation, regulation or co-occurrence, depending on the context.}
Several representations of temporal networks exists, each associated to different algorithms and methods. 
For example, as a sequence of static graphs representing time windows over which the activity is aggregated\cite{Mucha2010},
as contact sequences when events are instantaneous in continuous time
or as interval graphs\cite{Holme2012} or link streams\cite{Sun2007,Latapy2018} in continuous time with events that may have a duration.
The study of the dynamics and structure of time-dependent networks 
has attracted many contributions from several fields such as 
sociology\cite{Stadtfeld2017,Butts2008,Moody2005}, computer 
science\cite{Sun2007,Takaffoli2011,Rossetti2017,Folino2014,Aynaud2011,Viard2016}, 
epidemiology\cite{Holme2014,Valdano2015}, mathematics and network 
science\cite{Mucha2010,Peixoto2017,Aslak2018,Petri2014,Palla2007,Matias2017,
Ghasemian2016} (references are not exhaustive).

\rev{Community detection in networks is the task of extracting a simplified view of a network's structure and is fundamental to help understand the functioning of the systems they represent\cite{Fortunato2010}.
Loosely speaking, a community is a relatively dense sub-graph, and it may be called a module or a cluster depending on the field of application.
Within a temporal setting,}
Rossetti and Cazabet\cite{Rossetti2018}
classify dynamic community detection methods based on 
how \rev{the } dynamic communities \rev{they find depend on time} in three categories ranked in increasing degree of their temporal smoothness: 1) Instant optimal, when the community structure at time $t$ depends only on the topology of the network at 
that time (e.g. \cite{Palla2007,Takaffoli2011}); 2) Temporal Trade-off, when the 
community structure at time $t$ depends on the topology of the network at $t$ 
but also on the past topology or past community structure (e.g. 
\cite{Rossetti2017,Folino2014}); 3) Cross-Time, when the community structure at 
time $t$ depends on the entire network evolution (e.g. 
\cite{Matias2017,Ghasemian2016,Viard2016}).

Critically, most methods aggregate the temporal 
dimension over a sequence of time windows, transforming the network in a 
sequence of static networks defined on a discrete time grid,
\rev{hence losing the precise ordering of the edge activations within each slice.
This is necessary as these approaches rely on a static concept of communities, i.e. defined as a group of nodes that are more densely connected with each other than with the rest of the network and, to be meaningful in a temporal context, the notion of density of connections necessarily implies connections considered over some time interval.}
\rev{They then} either apply standard community detection 
algorithms for static networks to each aggregated time 
slice \rev{and follow the evolution of the communities across time slices with special algorithms}\cite{Folino2014,Holme2015,Liechti2019} or consider each slice as a layer of a 
multilayer network and apply a community detection method to the entire 
multilayer network (e.g. \cite{Mucha2010,DeDomenico2015})\rev{, hence defining communities over extended periods of time}.
\rev{Methods based on an underlying dynamical process\cite{Rosvall2008,Delvenne2010}, taking place on each slice\cite{Aynaud2010,guo2014evolutionary} or on the entire multilayer network\cite{Mucha2010,DeDomenico2015}, consider a process decoupled from the intrinsic time of the system under study in order to guarantee its stationarity.}
Statistical approaches have also 
been developed, for example Peixoto and Rosvall\cite{Peixoto2017} have 
generalized the framework of stochastic block model inference to a dynamical 
framework by including a Markov chain in the inferred model. Their generative 
model approach takes into account continuous time Markov chain and 
\rev{can capture the ordering of events, however it requires that the Markov chains describing the system must be stationary on different epochs.}

\label{rev:methoddescr}
Here, we propose a novel method that considers random walks (RW) evolving on the network and restricted by the activation times of the edges.
\rev{We consider the similarity of diffusion patterns over a given time interval as a way to cluster nodes together without resorting to temporal aggregation and while only considering time respecting paths.
This approach generalizes the notion of cluster density used in static methods, such as Markov stability\cite{Delvenne2010}, to the temporal case.
We derive quality functions that allows one to find partitions
that best cluster the flow of random walkers and that do not need to be evaluated using the stationary state of the diffusion process. This is necessary as the existence of such a state is not guaranteed when considering a process evolving with the temporal network.}
We show that the temporal evolution of networks
leads to potentially asymmetrical relations between vertices that
can be captured by using two network partitions
for a given time interval: the \emph{forward partition}
and the \emph{backward partition} \rev{that cluster nodes from the point of view of the beginning and end of the time interval, respectively}.
\rev{We leverage the novel possibility of our method to be used with non-stationary realizations of a diffusion process to find dynamic communities relating the temporal influence between a small group of nodes and the entire network}.
We also show that our method allows one to reveal different dynamical scales
present in temporal networks by \rev{using a RW process evolving with the network and varying its rate of diffusion.} 
\rev{When compared to methods that necessitate to aggregate the network evolution in several static time windows, we find that our method can captures dynamical scales existing at rates that are lost in the aggregation procedure.}
Our framework generalizes the concept of Markov stability\cite{Delvenne2010,Lambiotte2014}
and dynamical embeddings\cite{Schaub2019} to  
the case of temporal networks \rev{without having to be evaluated at stationarity}.

\section*{Results}

\label{sec:temp_net}

\subsection*{Temporal flow stability}

We consider the general case of a temporal network with a set of $N$ vertices $V$,
a set of $M$ events $E$ and 
two sets of $M$ not necessarily distinct starting and ending times, $T^s$ and $T^e$.
Here, the term \emph{event} is used to represent the generalization of edges to the temporal case\cite{Holme2019}.
Event $i$ can be written as a tuple $e_i = (u, v, t^s_i, t^e_i)$ where $u$ 
and $v$ are the source and target vertices, respectively, $t^s_i$ is the
time at which the edge becomes active and $t^e_i$ is time at which the event ends, with $t^e_i\ge t^s_i$.
This definition is equivalent to the ones of interval graphs\cite{Holme2012} or link streams\cite{Latapy2018}
and can also be used to describe more restrictive definitions of temporal networks
with instantaneous events or as sequences of static graphs.
\label{rev:streamgraphs}
\rev{A more general model of temporal networks, the stream graph model\cite{Latapy2018}, also takes into account nodes with specific activation times. Our framework does not distinguish nodes that are absent from nodes that are present but inactive.}
We want to find a partition of the 
network in $c$ non-overlapping communities that describes well its structure.
The $N\times c$ indicator matrix, $\matr{H}$, records which vertex belongs to which community, 
e.g. each row of $\matr{H}$ is all zeros except for a one indicating the cluster to which the vertex belongs.

We consider a random walk (RW) process starting on all nodes of the network at time $t_1$ with a density probability described by the $1\times N$ row-vector $\vect{p}(t_1)$ and ending at $t_2$ ($t_1<t_2$) with a density $\vect{p}(t_2)$. The random walk evolution is restricted by the activation of the network's edges and the transition probability matrix of the RW, $\Tmat(t_1,t_2)$, is such that $\pvec(t_2)=\pvec(t_1)\Tmat(t_1,t_2)$ (see Methods and Materials). 
\rev{Random walks are at the core of a variety of methods for community detection on static networks. However, their direct application to a temporal setting does not necessarily provide a satisfying answer.
As an illustration, consider the} framework of Markov Stability \rev{which clusters a} network in groups of nodes where the random walkers are likely to remain for a given time. This can be achieved by clustering the covariance matrix of the process which encode probabilities for walkers to start on a given node and end on another after a certain time minus the same probability for independent walkers\cite{Delvenne2010}.
For a general, not necessarily stationary, random walk on a temporal network, the $N \times N$ covariance matrix between $t_1$ and $t_2$ is given by (see Methods and Materials)
\begin{equation}
    \matr{S}(t_1, t_2) = \matr{P}(t_1) \matr{T}(t_1, t_2) - \vect{p}(t_1)^\mathsf{T} \vect{p}(t_2),
    \label{eq:autocov} 
\end{equation}
where $\matr{P}(t_1)=\textrm{diag}(\pvec(t_1))$.
In the case of static networks, \rev{and taking $\pvec(t_1)=\pvec(t_2)=\vect{\pi}$ to be the stationary distribution of the random-walk process (which is defined if the graph is strongly connected),} this expression reduces to 
the framework of Markov stability\cite{Delvenne2010,Lambiotte2014}.

 \rev{In a temporal setting,} a stationary state does not necessarily exist
\rev{and it is in general ill-defined in the case of a network with a finite time window. For this reason, the initial distribution is not uniquely defined and we argue that it can be chosen by the user depending on its purposes.}
\rev{This} framework provides the ground to detect relevant multi-scale structures in temporal networks and opens the door for a more general understanding of clustering in networks \rev{with non-stationary processes}.
However, constructing a quality function using eq. (\ref{eq:autocov}) directly
does not satisfyingly solve the temporal community detection problem and  this quality function needs some slight, yet conceptually important, modification. 

\rev{To show so, we focus} on the case of temporal networks with undirected events. 
Interestingly, whether the events of the temporal networks have a direction or not,
the transition matrix of the random walk between two times is, in general, asymmetric.
Indeed, the time ordering of events can result in different probabilities for 
going from a particular node $i$ at $t_1$ to a node $j$ at $t_2$ than going from $j$
at $t_1$ to $i$ at $t_2$\cite{Scholtes2014}, even if each event allows walkers to 
travel in both directions.
As a consequence the covariance matrix $\matr{S}(t_1,t_2)$ is also asymmetric in general.
For temporal network, the concept of community needs to take into account the temporal evolution of the network
and the temporal asymmetry potentially arising from it.
The element $(i,j)$ of the covariance $\Smat(t_1,t_2)$ (eq. \ref{eq:autocov}) gives the probability 
that a walker is on node $i$ at $t_1$ and on node $j$ at $t_2$ minus the 
same probability for two independent walkers. 
Directly clustering $\Smat(t_1,t_2)$ in diagonal blocks would force
a symmetric relation between nodes based on the RW state at two different times, as
rows of $\Smat(t_1,t_2)$ refer to the state in $t_1$ and 
columns of $\Smat(t_1,t_2)$ to the state in $t_2$.
By construction, $\Smat(t_1,t_2)$ considers the positions of the random walker at different times, 
and thus builds communities across time that are not synchronous, i.e. that aggregate nodes by comparing their states at different times. 

\begin{figure}[htp]
\centering
 \includegraphics[width=0.9\linewidth]{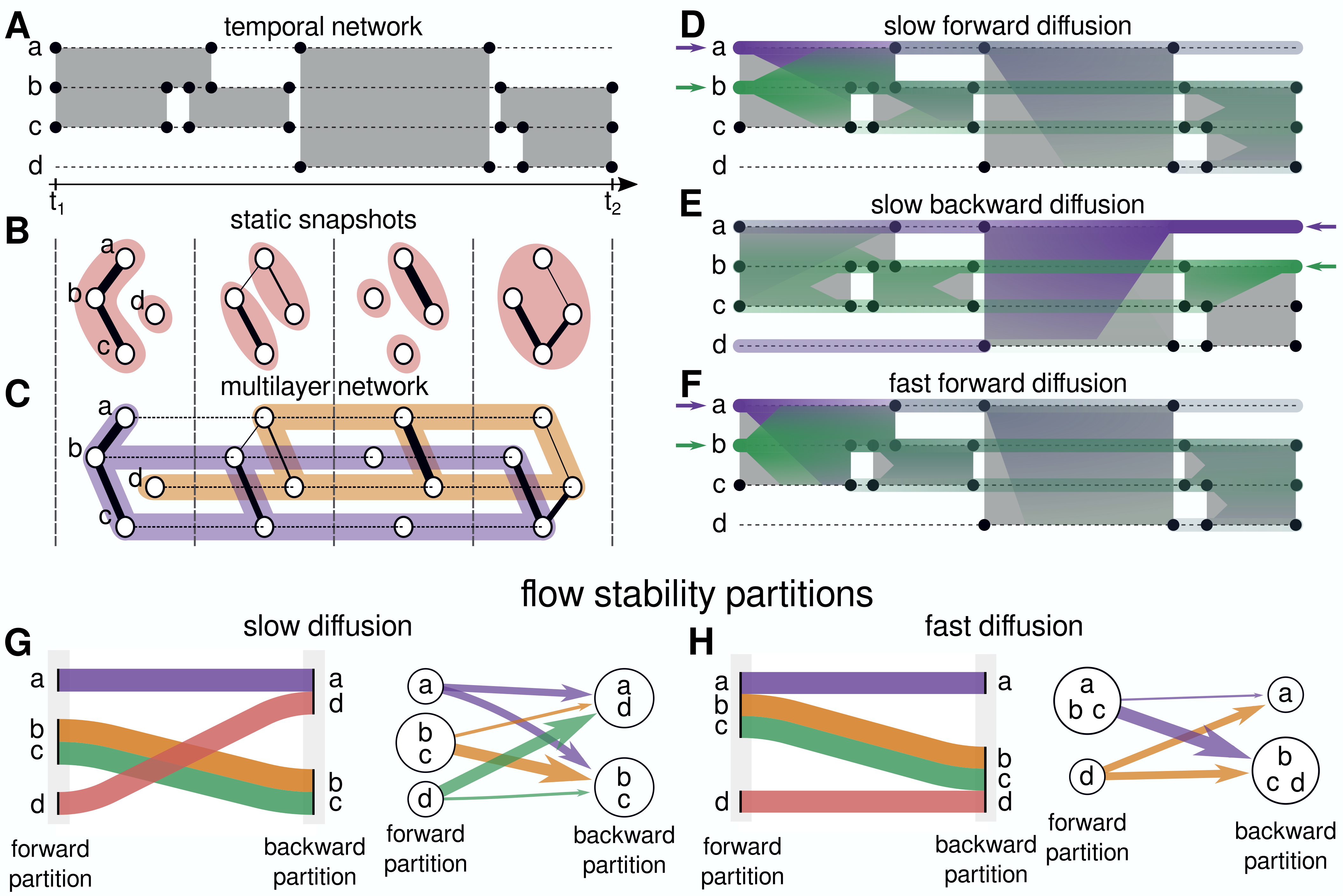}
 \caption{\rev{\textbf{Schematic representation of the flow stability compared to other temporal community detection methods.}
 (\textbf{A}) Example of a temporal network with 4 nodes ($a, b, c$ \& $d$) and events joining pairs of nodes for different durations in continuous time.
 (\textbf{B}) A representation of the temporal network as four static networks each representing the aggregated edge activity over the corresponding time window. The communities found by modularity optimization are overlayed.
 (\textbf{C}) Multilayer representation of the sequence of static networks with inter-layer links added in-between layers.
 The communities found by optimizing the multilayer modularity\cite{Mucha2010} (in orange and purple) extend over several layers.
 (\textbf{D}, \textbf{E} \& \textbf{F}) Schematic representation of a diffusion process starting on nodes \revv{$a$ (purple) and $b$ (green)} in forward time (\textbf{D}) and backward time (\textbf{E}).
 The probability density of the process at time $t$, computed with the transition probability matrix $\Tmat(t_1,t)$ for the forward time and $\Tmat_{rev}(t_2,t)$ for the backward time, is represented by the transparency of the colors.
 Using a faster diffusion rate (\textbf{F}), the process explores larger areas faster and \revv{earlier} events become more important.
 \revv{The flow stability method groups nodes together based on the similarity of the diffusion processes starting on them and takes into account the ordering of the events and the dynamics of the network}.
 (\textbf{G}) Two representations of the forward and backward partitions found with the flow stability method: an alluvial diagram (left) showing how nodes move between communities and a graph (right) showing the communities as meta nodes and the transition probabilities of the random walk process between them as directed edges.
 \revv{Even by varying the aggregation window length, the initial and final static partitions cannot fully reproduce the results of the flow stability (see Tab. S\ref{tab:comparison_slice_len}).}
 (\textbf{H}) Using a faster diffusion rate, nodes \revv{$a$} and \revv{$b$} are clustered together in the forward partition.}}
 \label{fig:schema}
\end{figure}

To make the similarity between the nodes synchronous and, concurrently, to capture the network evolution from $t_1$ to $t_2$,
we propose to consider two partitions, effectively clustering the 
rows and columns of covariances separately, 
and grouping together
nodes based on their simultaneous state time and on the forward or backward evolution 
of the RW process (see Supplementary Text \nameref{sec:co-clust}).
This idea builds on the concept of dynamical embeddings of network\cite{Schaub2019} but generalized to temporal networks.
We consider that two nodes are in the same \emph{forward} community if 
the random walkers starting on them at $t_1$ tend to stay on the same nodes during the evolution of the network until $t_2$.
To capture the temporal asymmetry, we also consider \emph{backward} communities. 
A first possibility is to define \emph{backward} communities by considering the random process that started at $t_1$ and saying that two nodes are in the same \emph{backward} community if the
random walkers that end on them at $t_2$ tended to stay on the same nodes from $t_1$ to $t_2$.
A second possibility is to consider the reverse evolution of the network, where
random walkers start at $t_2$ and diffuse until $t_1$. 
In this case, the \emph{backward} communities are defined as the \emph{forward} 
communities, but by reversing the direction of time.
\rev{Figure \ref{fig:schema} illustrates the concept of the flow stability method on a simple example and compares it with other temporal community detection methods}.

To find nodes from which random walkers tend to end up on the same node, we
consider the process following the evolution of the network
from $t_1$ to $t>t_1$ and followed by the inverse process going from $t$ to $t_1$.
The transition probability matrix corresponding to the inverse process, defined as 
the matrix $\matr{T}^{\textrm{inv}}(t,t_1)$ satisfying $\vect{p}(t)\matr{T}^{\textrm{inv}}(t,t_1)=\vect{p}(t_1)$, is given by Bayes' theorem as
$
\matr{T}^{\textrm{inv}}(t,t_1) = \matr{P}(t)^{-1}\matr{T}(t_1,t)^\mathsf{T}\matr{P}(t_1)
$\cite{Perez-Nimo2018},
where $\vect{p}(t)=\vect{p}(t_1)\matr{T}(t_1, t)$.
\rev{The $\Tmat^{\textrm{inv}}(t,t_1)$ matrix encodes the transitions probabilities to go from a state $\pvec(t)$ of a specific process back to the initial condition of this same process $\pvec(t_1)$, e.g. going backward in time in Fig.~\ref{fig:schema}D or F}.\label{rev:diffusion_schema}
The corresponding covariance is 
\begin{align}
\matr{S}_{\text{forw}}(t_1,t) &= \matr{P}(t_1)\matr{T}(t_1,t)\matr{T}^{\textrm{inv}}(t,t_1) - \vect{p}(t_1)^\mathsf{T}\vect{p}(t_1) \notag\\
&= \matr{P}(t_1)\matr{T}(t_1,t)\matr{P}(t)^{-1}\matr{T}(t_1,t)^\mathsf{T}\matr{P}(t_1) - \vect{p}(t_1)^\mathsf{T}\vect{p}(t_1),
  \label{eq:autocov_forw}
\end{align}
which is symmetric by construction and has element $(i,j)$ giving the probability that two random walkers starting in $i$ and $j$ at $t_1$ 
finish on the same node at $t$ minus the probability that two independent walkers start in $i$ and $j$ at $t_1$.
The matrix $\matr{S}_{\text{forw}}(t_1,t)$ contains the product of $\Tmat(t_1,t)$ and $\Tmat(t_1,t)^\mathsf{T}$
and can be seen as a matrix measuring the similarity of the rows of $\matr{T}(t_1,t)$.
Indeed, our method can be seen as a way to perform a co-clustering of the transition matrix (see Supplementary Text \nameref{sec:co-clust}).
Moreover, this matrix is properly normalized, i.e. each rows and columns sum to zero, which is necessary for 
optimization method such as the Louvain algorithm\cite{Blondel2008}.

Similarly, we can define a backward process by reversing time which results in the following covariance matrix:
\begin{align}
 \matr{S}_\text{back}(t_2,t) &= \matr{P}(t_2)\matr{T}_\textrm{rev}(t_2,t)\matr{T}^\textrm{inv}_\textrm{rev}(t,t_2) - \vect{p}(t_2)^\mathsf{T}\vect{p}(t_2) \notag\\
 &= \matr{P}(t_2)\matr{T}_\textrm{rev}(t_2,t)\matr{P}(t)^{-1}\matr{T}_\textrm{rev}(t_2,t)^\mathsf{T}\matr{P}(t_2) - \vect{p}(t_2)^\mathsf{T}\vect{p}(t_2),
 \label{eq:autocov_backw}
\end{align}
whose element $(i,j)$ gives the probability that two random walkers 
starting in $i$ and $j$ at $t_2$ and following the reversed evolution of the network
finish on the same node at $t_1$ minus the probability that two independent walkers start in $i$ and $j$ at $t_2$.
Here, $\Tmat_\textrm{rev}(t_2,t)$ is computed as $\Tmat(t_1,t)$, but by considering the reversed evolution of the network since $t<t_2$ (see Methods and Materials).
\rev{Figure \ref{fig:schema}E shows and example of this backward diffusion process.
Similarly to the forward case, $\Tmat^\textrm{inv}_\textrm{rev}(t,t_2)$ is given by Bayes' theorem and encodes the transition probabilities to go from a state $\pvec(t)$ of a specific backward process back to the initial condition, $\pvec(t_2)$, of this process. This can be seen as going forward in time in Fig.~\ref{fig:schema}E.}

In this study, we consider 
$\Smat_\text{forw}(t_1,t)$ and $\Smat_\text{back}(t_2,t)$ for $t_1<t<t_2$ with two corresponding initial conditions $\vect{p}(t_1)$ and $\vect{p}(t_2)$ 
taken as uniform distributions over all nodes, i.e. the maximum entropy distribution, for the general study of the dynamics of a temporal network between $t_1$ and $t_2$.
This allows one to consider the forward and backward partitions independently as they both depend on their own process.
We also investigate an example of clustering of a specific random process defined by a non-uniform initial probability distribution (see section \nameref{sec:aps}).
\rev{We discuss in Supplementary Text \nameref{sec:inv_covar} an alternative definition of the backward covariance based on the same process than for the forward covariance.}

We define the \emph{forward} and \emph{backward flow stability functions} as 
\begin{equation}
 I^\textrm{flow}_\textrm{forw}(t_1,t_2;\matr{H}_{f}) = \rev{\frac{1}{t_2-t_1}}
 \textrm{trace}\left[\matr{H}_{f}^\mathsf{T}\int_{t_1}^{t_2}\matr{S}_\text{forw}(t_1, t)dt\matr{H}_{f}\right]
\label{eq:forw_int_stability}
\end{equation}
and
\begin{equation}
 I^\textrm{flow}_\textrm{back}(t_1,t_2;\matr{H}_{b}) = \rev{\frac{1}{t_2-t_1}}
 \textrm{trace}\left[\matr{H}_{b}^\mathsf{T}\int_{t_2}^{t_1}\matr{S}_\text{back}(t_2, t)dt\matr{H}_{b}\right].
\label{eq:back_int_stability}
\end{equation}

The two partitions that maximize the forward and the backward flow stability functions, described 
by $\matr{H}_f$ and $\matr{H}_b$ respectively, 
describe the temporal evolution of the network structure between $t_1$ and $t_2$.
By taking the integral of the covariance over $t$, we find the most persistent communities
during the entire time interval and give more weight to early times for the forward stability, 
or late times in the case of the backward stability, assuring that the time ordering of events is 
captured by both partitions. 
The integration correctly captures the time ordering even when a different ordering of the events results in the same final transition matrix \rev{, i.e. when} inter-event transition matrices commute.
\label{rev:timeweighting}
\rev{The weight of early times compared to later time in the forward partition, or late times compared to early times in the backward partition, can be controlled by varying the rate of the random walk process. 
We illustrate this effect with an analytic example in the Supplementary Text \nameref{sec:time-weighting} and Supplementary Fig. S\ref{fig:time_weighting}.
However, our method gives a simplified description of the entire evolution of a network during a time interval with only two partitions and from the point of view of the starting and ending times of the interval. 
Details about the structure and dynamics in the middle of the interval may therefore be lost in the coarse graining procedure.
When details about the dynamics happening in the middle of the interval are wanted, the time interval can be divided in a series of time windows for each of which two partitions are computed.
In this case, compared to other methods that represent a temporal network as a sequence of static aggregated time windows (see Fig. \ref{fig:schema}B \& C), our approach has the advantage of preserving information about the dynamics inside each time window.
We give an example of such an approach in the section \nameref{sec:mice}.
We also give results of the flow stability clustering applied on typical dynamic community events in Supplementary Fig. S\ref{fig:community_events}.\label{rev:community_events}}

\subsection*{\rev{Example of temporal network with asymmetric temporal paths}}
\label{sec:trans_toy}

\begin{figure}[ht]
\centering
 \includegraphics[width=\linewidth]{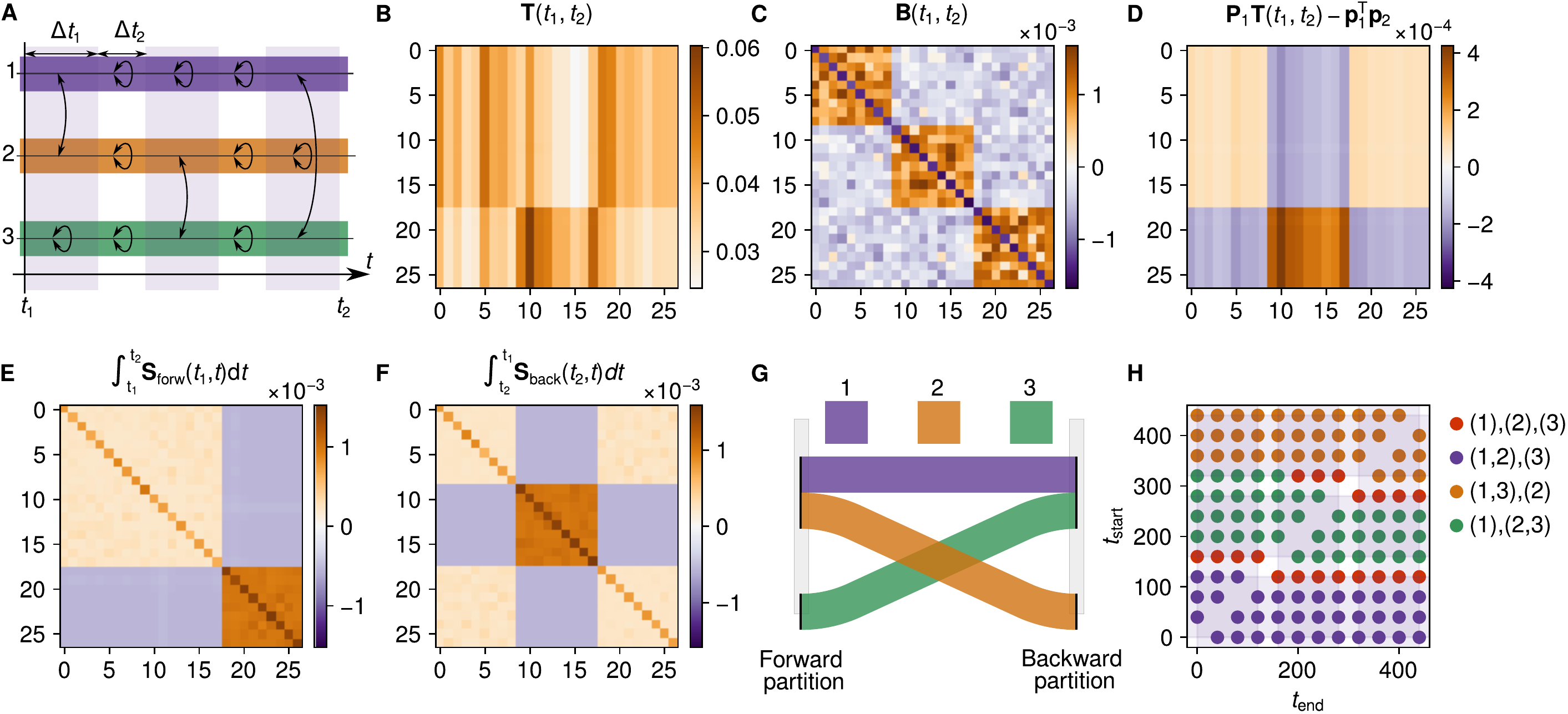}
 \caption{\textbf{Flow clustering of a synthetic network with \rev{asymmetric temporal paths}}.
 (\textbf{A}) Representation of the different phases of interactions between the three groups of nodes.
 (\textbf{B}) Random walk transition matrix computed between the start, $t_1$, and the end of the three phases, $t_2$.
 (\textbf{C}) Modularity matrix computed on the aggregated network ($\matr{B}=\matr{A}-\frac{\matr{k}^\mathsf{T}\matr{k}}{2m}$, where $\matr{A}$ is the aggregated adjacency matrix, $\matr{k}=\matr{A}\matr{1}$ and $2m=\sum_i k_i$).
 (\textbf{D}) Covariance matrix, $\matr{S}(t_1,t_2)$, of the random walk process defined in eq. (\ref{eq:clustered_autocov}).
 (\textbf{E}) Integral of the forward process covariance matrix (eq. (\ref{eq:autocov_forw})).
 (\textbf{F}) Integral of the backward process covariance matrix (eq. (\ref{eq:autocov_backw})).
 (\textbf{G}) Representation of the forward and backward process in an alluvial diagram.
 (\textbf{H}) Best partitions found when varying the starting and ending times of the considered interval.}
 \label{fig:trans_covar_matrices}
\end{figure}

As a simple model of temporal network where 
\rev{the time ordering of events leads to relations between nodes that could not be captured by a temporal aggregation in a static network},
we consider the following network made of three groups of nine vertices each.
Vertices are activated at random times drawn from an exponential distribution with parameter 
$\lambda_\textrm{activ}$ (Poisson process). 
When a vertex is activated it chooses another vertex according to a certain rule and the duration
of the interaction is drawn from an other exponential distribution with parameter $\lambda_\textrm{inter}$.
The system follows two types of successive interactions: 
I1) during $\Delta t_1$ the vertices of two of the groups interact with one another with probability $p_1>1/2$ 
and with any other vertices in the network with probability $1-p_1$ while the vertices of the third group only interact
with each other;
I2) during $\Delta t_2$ each vertex interacts 
with other vertices of its group with a probability $p_2> 1/2$ and with any vertices in the network
with probability $1-p_2$.
We generate a realization of the temporal network by running a simulation composed of three phases of interactions
I1 separated by I2 phases as shown in Fig. \ref{fig:trans_covar_matrices}A).
During the first I1 phase, groups one and two interact, during the second I1 phase, groups two and three interact and 
finally during the third I1 phase groups one and three interact.
\rev{If it were not for the small probability to reach any node in the network (if $p_1=p_2=1$), the temporal paths in this network would not all be transitive, i.e. the existence of time respecting paths from a node $i$ to a node $j$ and from node $j$ to $k$ would not guarantee the existence of a time respecting paths from node $i$ to $k$}.
\rev{With $p_1<1$ and $p_2<1$, the situation is less dramatic, but} the ordering of interactions creates temporal paths with asymmetric probabilities: for example there are many paths that start in group one, are in group two at the end of the first I1 phase and in group three at the end of the second I1 phase. However, there are almost no paths starting from group three, going to group two and group one during the same time lapse.
Defining communities in this temporal network is not straightforward. 
If we were to discard the temporal dimension, we would find that nodes are more densely connected with other nodes of the same group, however the temporal pattern of interactions between groups would be lost.
A good temporal partition in communities should offer a simplified description of the network structure and
its evolution. In this case, the three groups should be identified as well as the ordering of their interactions.
We show that we are able to achieve this by defining communities in terms of the flow of random walkers restricted
by the edges activations.
We run a simulation with the following parameters: $\lambda_\textrm{activ}=1$, $\lambda_\textrm{inter}=1$, 
$p_1=0.95$, $p_2=0.95$, $\Delta t_1=120$ and $\Delta t_2=40$.

Figure \ref{fig:trans_covar_matrices}B shows the transition matrix, $\matr{T}(t_1,t_2)$
computed from the resulting realization of the temporal network, between the start and the end of the three phases, 
using a continuous time random walk model (see Materials and Methods) and 
Fig. \ref{fig:trans_covar_matrices}C shows the modularity matrix obtained when aggregating
the temporal dimension. 
As expected, when aggregating the temporal activity, the temporal pattern of interaction is lost and only the
three groups are visible.
Figure \ref{fig:trans_covar_matrices}D shows the covariance matrix $\matr{S}(t_1,t_2)$ (eq. \ref{eq:clustered_autocov}).
The temporal asymmetry of the system evolution is captured by the asymmetry of $\matr{S}(t_1,t_2)$.
The two symmetric matrices corresponding to the forward and backward integrals of the symmetrized covariances (eqs. \ref{eq:autocov_forw} and \ref{eq:autocov_backw}, respectively) are shown in Figs. \ref{fig:trans_covar_matrices}E \& F. They capture the similarities between the rows and columns of $\matr{S}(t_1,t_2)$, integrated over the entire system evolution.
The partitions that best describe them are found by optimizing the forward and backward flow stability functions (eqs. \ref{eq:forw_int_stability} and \ref{eq:back_int_stability}, respectively) and are represented in Fig. \ref{fig:trans_covar_matrices}G in an alluvial diagram\cite{Lupton2017}.
The forward and back partitions, and the relation between them, capture the three groups and the fact that groups one and two interact together at the beginning, groups one and three interact together at the end, while
groups two and three ``exchange'' their position with group one during the evolution of the network.
Figure \ref{fig:trans_covar_matrices}H shows the best partition found with our method 
by varying the starting and ending times of the considered interval.
When $t_{start} < t_{end}$ (below the diagonal), the best partition is the forward partition (eq. \ref{eq:forw_int_stability})
and when $t_{start} > t_{end}$ (above the diagonal), the best partition is the backward partition (eq. \ref{eq:back_int_stability}).
The alluvial diagram in Fig. \ref{fig:trans_covar_matrices}G, capture the global structure
and dynamic of the system during its entire evolution, while Fig. \ref{fig:trans_covar_matrices}H 
allows to reveal the detailed timing of the interactions between groups.

\subsection*{Temporal multiscale community detection}

An important point concerning community detection methods based on the optimization of a quality function,
such as modularity optimization, is that the quality function implicitly restrict the size of the communities 
maximizing it\cite{Fortunato2007}.
\rev{Quality functions including an explicit resolution parameter permit to overcome this problem.}
For example, the time parameter of the Markov stability framework serves as a resolution parameter that 
generalizes the modularity\cite{Delvenne2010,Lambiotte2014} and allows to find communities at all scales in the network\cite{Schaub2012}.
In the case of temporal network, the concept of scale must take into account both the speed at which 
the network changes and the different sizes of its structures.

Here, the rate at which random walkers jump from nodes to nodes serves as a natural resolution parameter
that controls how far walkers moves during a certain time window.
We use a continuous time random walk on networks, 
which can also be described as a continuous time Markov chain\cite{Seneta1981,Masuda2016}, 
and we assume that, when an edge is active, walkers have a constant probability of jumping per unit of time given by the rate $\lambda$,
or equivalently an average waiting time $\tau_w = 1/\lambda$, even if the topology of the network is changing
with time.
The transition matrix therefore depends on the evolution of the network and on the random walk waiting time, i.e. we have $\matr{T}(t_1,t_2)=\matr{T}(t_1,t_2;\tau_w)$.
In the case of a temporal network where all edges are constant in time, i.e. a homogeneous Markov Chain, we have

\begin{equation}
\matr{T}(t_1,t_2;\tau_w)=e^{-\frac{t_2-t_1}{\tau_w}\matr{L}} 
=  \matr{I}+\sum_{n=1}^{\infty}\frac{\left(-\lambda^\star\matr{L}\right)^n}{n!}, 
\label{eq:trans_mat}
\end{equation}
where $\matr{L}=\matr{I}-\matr{D}^{-1}\matr{A}$ 
is the random walk Laplacian of the network\cite{Masuda2016}
and $\lambda^\star =\frac{t_2-t_1}{\tau_w}$ is a normalized random walk rate.
When computing the adjacency matrix $\matr{A}$, we add self-loops on isolated nodes in order to keep the transition matrix stochastic.
In the case where the network topology is changing in time, the transition matrix is computed 
as the time-respecting product of inter-event transition matrices (see the details in the Materials and Methods section).
We observe that varying $\tau_w$ in eq. (\ref{eq:trans_mat}) allows to ``zoom'' in or out on the network.
For $\lambda^\star = 0$ (or $\tau_w \rightarrow \infty$), i.e. for extremely slow walkers, 
$\matr{T}(t_1,t_2;\tau_w)=\matr{I}$ and walkers simply stay on their current nodes.
When $\lambda^\star = 1$ ($\tau_w = t_2-t_1$), on average walkers will have had the time to only jump to their direct neighbors.
For $\lambda^\star \gg 1$ ($\tau_w \ll t_2-t_1)$, i.e. for very fast walkers, the walker will have
explored their entire reachable surroundings and, unless the random walk is periodic, reached stationarity.

\begin{figure}[htbp]
\centering
 \includegraphics[width=\linewidth]{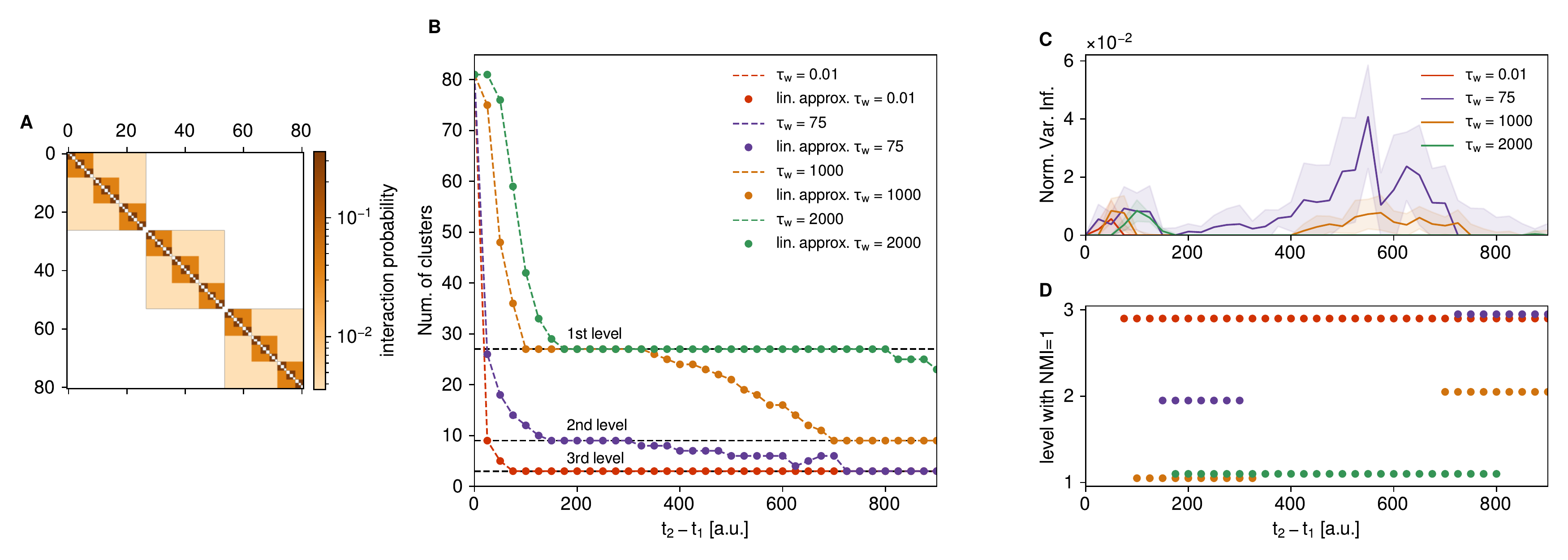}
 \caption{\textbf{Multiscale temporal network clustering with the flow stability}.
 (\textbf{A}) Interaction probabilities of the synthetic temporal block model showing
 a hierarchical structure with three levels.
 (\textbf{B}) Number of clusters as a function of the time interval length for different values of the 
 characteristic waiting time, $\tau_w$.
 (\textbf{C}) \rev{Average} normalized variation of information \rev{(NVI)} as a function of the time interval duration.
 \rev{Minima of the NVI indicate when the solutions of the optimization are robust. The filled band show the standard deviation computed across the 10 simulations.}
 (\textbf{D}) Correspondence between the optimal partitions found and the different levels
 of the network hierarchical structure}.
 \label{fig:hiera_clust}
\end{figure}

Figure \ref{fig:hiera_clust} shows an example of the multiscale detection capabilities of our method and a comparison
of the results obtained using the matrix exponential formulation to compute the transition matrices (eq. \ref{eq:trans_mat})
with the linearized version (eq. \ref{eq:lin_trans}).
For this example, we modeled temporal networks with 81 nodes using the same principle than for previous example with parameters $\lambda_\textrm{activ}=1/10$ and $\lambda_\textrm{inter}=1/10$.
At each activation time, a node select another nodes to interact with given by a different probabilities.
The interaction probabilities, shown in Fig. \ref{fig:hiera_clust}A, define a hierarchical structure with a first level of 27 groups of three nodes,  
a second level with 9 groups of 9 nodes and a third level with 3 groups of 27 nodes. 
We choose the interaction probabilities such that $p_1/p_2=10$, $p_1/p_3=100$ and $p_4=0$, where $p_1$ is the probability
of a node to interact with nodes of the same first level group, $p_2$ is the probability
to interact with a node of the same second level group, $p_3$ is the probability to interact with
with nodes of the same third level and $p_4$ is the probability to interact with any other nodes.
Figure \ref{fig:hiera_clust}B displays the number of communities found by our method, 
with the computation using the matrix exponential
and the linear approximation, for different 
values of the average random walk waiting time ($\tau_w$) as a function of the time interval considered.
We run 10 simulations and display the number of communities of the most common optimal partition found 
among the 10 simulations.
To find the optimal partition, we run the Louvain algorithm\cite{Blondel2008} 50 times for each simulation
and keep the partition maximizing the forward integral flow stability (eq. \ref{eq:forw_int_stability}).
In this case, the network evolution is stationary and therefore using the backward integral flow stability
gives similar results.
The normalized variation of information (NVI) computed from the ensemble of partitions found by the Louvain algorithm
is shown in Fig. \ref{fig:hiera_clust}C. 
Minima in NVI indicate the intrinsic scales of the system\cite{Lambiotte2014} and therefore allows to 
choose the relevant resolution parameters, i.e. the random walk characteristic waiting times.
We observe that depending on the time interval considered,
we are able to recover the three scales of the system
using different combinations of the waiting time parameter and that they correspond 
to minima in NVI.
Figure \ref{fig:hiera_clust}D shows when the optimal partition found by our method corresponds
exactly to one of three levels partition, measured by the normalized mutual information (NMI).
For a given time interval, a slower random walk discovers the finer level (27 communities) while faster random walks discover the coarser levels (9 and 3 communities).
As time progresses random walks 
go from the first level to the second level (e.g. $\tau_w = 1000$ in Fig. \ref{fig:hiera_clust}) or from the second to the third level (e.g $\tau_w=75$ in Fig. \ref{fig:hiera_clust}), discovering coarser and coarser scales.
Figure \ref{fig:hiera_clust}B also shows that the linear approximation (circles) agrees very well
with the computation using the matrix exponential (dashed lines) and allows to detect the different scales similarly in both regimes 
of the approximation (eq. \ref{eq:lin_trans}).
Here, the average duration between changes in the network is $\simeq0.1$ time units, so on average, $\lambda^\star < 1$ for $\tau_w>0.1$ and $\lambda^\star > 1$ for $\tau_w<0.1$.

\begin{figure}[htbp]
\centering
 \includegraphics[width=\linewidth]{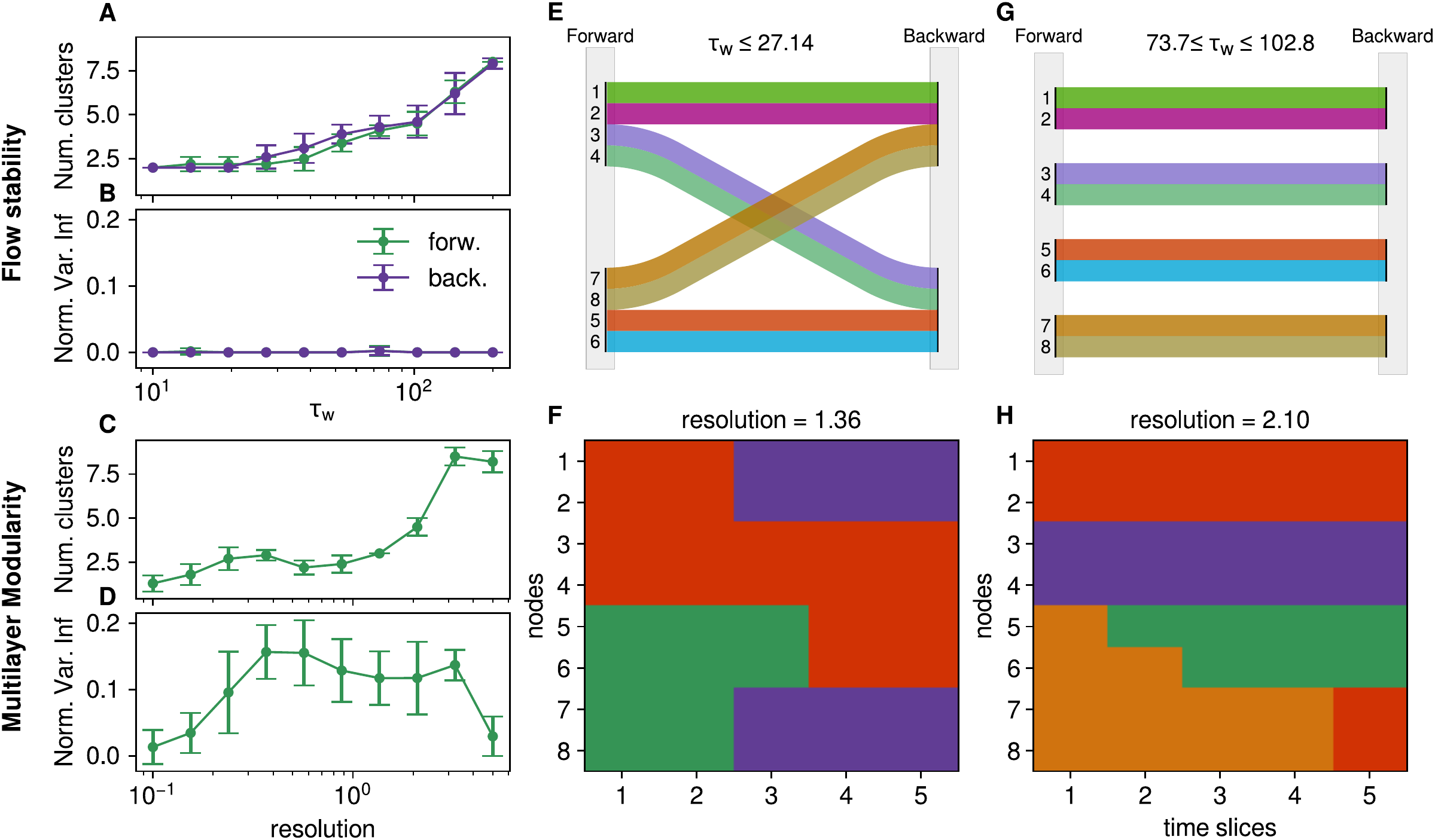}
 \caption{\textbf{Comparison of the flow stability method with the multilayer modularity on synthetic network with a continuously changing structure.}
 The connections of the eight nodes changes linearly from an initial structure in two communities: (1, 2, 3, 4) \& (5, 6, 7, 8), to a final structure: (1, 2, 7, 8) \& (3, 4, 5, 6).
 (\textbf{A} \& \textbf{C}) Average and standard deviation (SD) of the number of clusters as a function of the resolution parameter for the flow stability (FS) and multilayer modularity (MM), respectively.
 (\textbf{B} \& \textbf{D}) Average and SD  of the normalized variation of information (NVI) as a function of the resolution parameter for the FS and MM, respectively.
 (\textbf{E} \& \textbf{G}) Forward and backward partitions found with the flow stability for two different ranges of values of the resolution parameter that capture the two dynamic scales of the temporal network.
 (\textbf{F} \& \textbf{H}) Partitions found with the MM at two different values of the resolution parameter corresponding to local minima of the NVI.}
 \label{fig:ortho_cont_compa}
\end{figure}

As the random walk process is evolving with the temporal network, varying
the rate of the random walk can not only capture different co-existing structural scales 
but also capture \rev{dynamic} changes in \rev{structure}.
\rev{To demonstrate this, we run 10 simulations of our synthetic temporal network model with 8 nodes and interaction probabilities that change linearly from the structure in two communities (1, 2, 3, 4) \& (5, 6, 7, 8) at $t=0$ to a structure with the two communities (1, 2, 7, 8) \& (3, 4, 5, 6) at $t=100$.
Inside each communities, the interactions probabilities are uniform.
The activation rate of the nodes also change linearly between $t=0$ and $t=100$ from $\lambda_\textrm{activ}=1$ to $\lambda_\textrm{activ}=2$ for nodes 1, 2, 3 \& 4 and from $\lambda_\textrm{activ}=2$ to $\lambda_\textrm{activ}=1$ for nodes 5, 6, 7, \& 8. The event duration distribution is kept constant at $\lambda_\textrm{inter}=1$.
We apply the flow stability method over the entire time interval $t=0$ to $t=100$
and compare it with results obtained with the multilayer modularity\cite{Mucha2010} (optimized using the Leiden algorithm\cite{Traag2019,leidenalg}) applied to multilayer representations of the 10 networks with 5 layers containing the aggregated activity of the edges over 5 time windows. The interlayer coupling parameter is first fixed at 1/10 of the global average edge weight.
Figure \ref{fig:ortho_cont_compa}A \& C show the average and standard deviation (SD), taken across the 10 simulations, of the number of communities found by both methods as a function of the resolution parameter (i.e. the characteristic waiting time for the flow stability).
For both methods, 50 run of the optimization algorithm are performed and the partition with the largest value of the objective function is kept.
Figure \ref{fig:ortho_cont_compa}B \& D show the NVI of the 50 partitions as a function of the resolution.
The NVI measures the variation in the set of 50 partitions found by the algorithms at each resolution. The average and SD are again computed across the 10 simulations.
We see that the multilayer modularity show a high value of the average NVI and of its SD for non-trivial partitions, while the flow stability partitions have almost all an average NVI of 0 with SD of 0. Two points have very small non-zero values.
This indicates that the multilayer modularity has difficulties dealing with gradual changes and does not find consistent partitions when run several times on the same realization of the simulation. On the other hand, the flow stability show a very consistent results across all resolutions.
Figures \ref{fig:ortho_cont_compa}E, F, G \& H show partitions at two different resolutions for both methods.
For the flow stability, the partitions are also consistent across simulations, for $\tau_w=10$, the result displayed in Fig.\,\ref{fig:ortho_cont_compa}E is found for all simulations and correctly captures the large scale dynamic of the system. This solution stays the most frequent across simulations until $\tau_w=27.14$ where this forward partition is found in 8/10 of the simulations and the backward partition in 5/10.
The partitions shown in Fig. \ref{fig:ortho_cont_compa}G, capturing the small scale evolution, are found in 9/10 simulations for the forward partition, 8/10 for the backward, at $\tau_w=73.7$ and 6/10 simulations for the forward and backward partitions at $\tau_w=102.8$.
The partitions found with the multilayer modularity shown in Fig. \ref{fig:ortho_cont_compa}F \& H correspond to the two resolutions with similar average NVI that are a local minima of the NVI curve. While Fig. \ref{fig:ortho_cont_compa}F captures features of the evolution of the network structure, the multilayer modularity show a large variability over repeated run of the optimization (large NVI) and over different realizations of the simulation. The partition shown in Fig. \ref{fig:ortho_cont_compa}F is the most common among the different simulations and appears in 3/10 simulations. For the resolution shown in Fig. \ref{fig:ortho_cont_compa}H the 10 simulations result in 10 different optimal partitions (we show the one for the simulation that has the smallest NVI). 
Similar behaviors are observed by increasing or decreasing the number of time slices. When only two time slices are used, partitions with the initial and final configurations are found for the two slices, however still with a large NVI as the optimization hesitates between two configurations ((1, 2, 3, 4) in the first slice connected to (1, 2, 7, 8) in the second slice or (1, 2, 3, 4) connected to (3, 4, 5, 6)) in unequal proportions. In this case a solution with a smaller NVI gives the partition in four elongated communities similar to Fig. \ref{fig:ortho_cont_compa}G. 
Increasing or decreasing the interslice coupling weight also results in high NVI until the same structure in four constant communities is found for large values of the interslice coupling.
This example demonstrates that, without a priori knowledge of the real underlying dynamics, extracting the dynamic communities of continuously changing networks with multilayer methods is challenging.
On the other hand, the flow stability method can consistently uncover dynamical changes in the structure of temporal networks within a single interval by only varying one parameter, the RW waiting time. 
Methods like the multilayer modularity have more difficulties to find robust solutions and require to tune many parameters (resolution, number of slices and interslice coupling).}

\subsection*{Real world examples}

\subsubsection*{Primary school contact network}

As a first real world application of our method, we use the high-resolution measurements of face-to-face contact patterns recorded in a French primary school in the context of the sociopattern project\cite{Stehle2011}.
Face-to-face contacts between 232 children and 10 teachers were recorded during two days with the help of 
RFID devices, worn on the chests of participants, with a 20\,s resolution.
This dataset is well suited for validating temporal clustering method as the contacts are naturally
restricted by the separation in 5 grades with 2 classes per grade. Each class has an assigned room and an assigned teacher, however during morning, lunch and afternoon breaks, children mix in the playground or in the canteen.
As these common spaces do not have enough capacity to host all the students at the same time, only two or three classes have breaks at the same time, and lunches are taken in two consecutive turns\cite{Stehle2011}.

\begin{figure}[ht]
\centering
 \includegraphics[width=\linewidth]{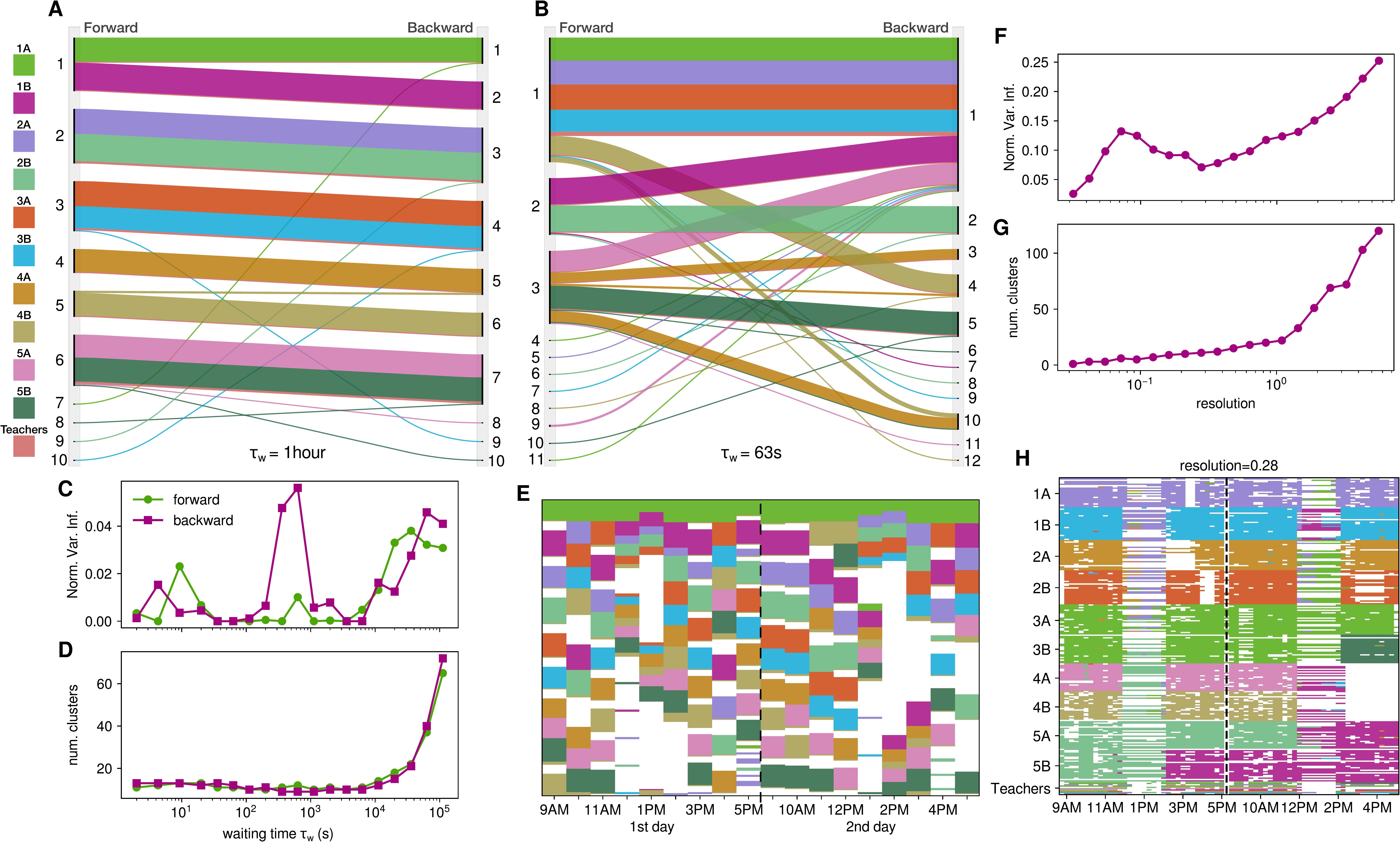}
 \caption{\textbf{Flow stability clustering of a face-to-face contacts in a primary school.}
 (\textbf{A} \& \textbf{B}) Alluvial diagram representing the forward and backward partitions at two 
 different scales corresponding to random walk rates of  $(1\,h)^{-1}$ and  $(63\,s)^{-1}$ respectively.
 (\textbf{C} \& \textbf{D}) Normalized variation of information and number of clusters
 of the best partition for different values of the characteristic waiting time.
 (\textbf{E}) Static clustering of hourly aggregated interactions using standard modularity
 optimisation.
 (\textbf{F} \& \textbf{G}) Normalized variation of information and number of clusters 
 found with the  generalized \rev{multilayer} modularity\cite{Mucha2010} as a function of the resolution parameter.
 (\textbf{H}) \rev{Multilayer} partition of the network corresponding to the minima in NVI in (\textbf{F}).}
 \label{fig:school_clust}
\end{figure}

We apply our method using the linear approximation of the transition matrices (eq. \ref{eq:lin_trans}) 
and perform 50 optimization of the forward and backward flow stability functions with the Louvain algorithm
for different RW characteristic waiting times.
The normalized variation of information (NVI) of the ensemble of partitions and the number of clusters of the best partition at each scale is shown in Figs. \ref{fig:school_clust}C \& D, respectively.
The NVI shows two minima, revealing the existence of two natural dynamical scales in the system, at $\tau_w= 63$\,s and $\tau_w= 1$\,h.
The forward and backward flow stability partitions corresponding to these two scales are shown in Figs. \ref{fig:school_clust}A \& B as alluvial diagrams.

The flow stability partitions found with a RW rate of $(1\,h)^{-1}$ (Fig. \ref{fig:school_clust}A) have 10 clusters
for the forward partition and 10 clusters for the backward partition that mostly group children of the same grades together but with some additional details.
Both partitions have also singleton clusters that correspond to children that were not present during the first day, for the forward partition, or second day, for the backward partition (see Tab. S\ref{tab:part_1h}).
Classes 1A and 1B are clustered together in the forward partition but separately in the backward partition, indicating that they spent less time together near the end of the time interval than near the beginning.
Classes 4A and 4B are separated in both the forward and backward partitions revealing that they spent less time together than other classes of the same grade.
All the other classes (2A, 2B, 3A, 3B, 5A and 5B) are clustered in pairs, per grade, in both the forward and backward partitions. 
Figure \ref{fig:school_clust}E shows the static clustering of hourly aggregated interactions using standard modularity optimization (with a resolution parameter corresponding to the minimum NVI taken over all hourly slices). 
\label{rev:validation}
Although this method removes all the temporal details within each hourly slice, it allows to coarsely represent the interactions between children during the two days
\rev{because the structures in this dataset changes according to the school hourly schedule. This hourly clustering allows to } \rev{verify the consistency} of the flow clustering \rev{obtained over the entire period}.
We see that, indeed, the classes 1A and 1B had lunch together during the first day, they are in the same static cluster at 12pm, 1pm and 2pm on the first day, but were separated during the lunch break of the second day.
We also see that classes 4A and 4B are separated during the morning and afternoon breaks of the first day and the morning and lunch breaks of the second day. 
In term of cumulative time of the contacts between all individuals of two different classes of the same grade, classes 4A and 4B are indeed the classes in the grade with the lowest cumulative contact time (439.3\,min) followed by classes 1A and 1B (582.7\,min) (see Tab. 3 in Ref.\cite{Stehle2011}).
All other grades have cumulative contact time between their classes above 966.7\,min.

Figure \ref{fig:school_clust}B shows the forward and backward partitions maximizing the flow stability with a RW rate of $(63\,s)^{-1}$ which capture changes happening at a faster scales than in Fig. \ref{fig:school_clust}A.
There are more forward and backward clusters with a small size than in Fig. \ref{fig:school_clust}A as they not only include children that missed the first or last day but also children, or small groups of children, that missed the morning of the first day or the afternoon of the second day (see Tab. S\ref{tab:part_63s}).
The largest forward cluster contains classes 1A, 2A, 3A and 4B. Fig. \ref{fig:school_clust}E shows that these classes are often together during breaks of the first day, in particular during the morning break of the first day. 
The backward partition contains a similar cluster with the addition of classes 1B \& 5A and without class 4B.
Indeed most of the children of class 4B leave after the lunch break of the second day, which is captured in the backward cluster 4 in Fig. \ref{fig:school_clust}B with an average last contact time of 12:52 PM, while the last contact time in cluster 1 is 05:07 PM (see Tab. S\ref{tab:part_63s}). 
Classes 1B and 5A join the largest cluster in the backward partition. 
Figure \ref{fig:school_clust}E shows that they are often clustered together during the second day and join the other
classes of the first cluster during the last hour.
We also see that, while class 4A is in cluster 3 with classes 5A and 5B in the forward partition, it is split in two separated clusters (3 and 10) in the backward partition.
\rev{Supplementary} Table S\ref{tab:part_63s} shows that the average last contact times for backward clusters 3 and 10 are 11:58 AM and 2:18 PM, respectively. 
The split in two clusters of class 4A is therefore due to the fact that a part of the class left before the lunch while the rest left after.

As a comparison to our method, we apply the generalization of the modularity to \rev{multilayer} networks developed in Ref. \cite{Mucha2010}.
We create network layers corresponding to an aggregation in windows of 15\,min with edge weights equal to the cumulative 
contact times during each time window. The interslice weight is set to the average edge weight across all layers.
Figures \ref{fig:school_clust}F \& G show the NVI and the number of clusters found by running the Leiden\cite{leidenalg} algorithm 50 times with the  generalized \rev{multilayer} modularity for each value of the resolution parameter.
Here, only one minima of the NVI is found and the corresponding partition is shown in Fig. \ref{fig:school_clust}H.
The partition captures the separation in grades and most of the separation in classes as well as some of the dynamics between classes.
The scale and resolution in this case does not include the concept of time, but consider the different layers as part of
a larger static network. 
In our method, the different scales correspond to different speed at which the network is traversed and the two partitions 
correspond to the different directions of the temporal evolution of the network.
We see that this allows us to discover two natural scales that describe the temporal network at two different levels: at the scale of 1\,h, we find the separation in different grades while at the scale of 63\,s we find a coarser scale describing the interactions in-between grades and classes.
\label{rev:multilaycomplex}
\rev{Community detection performed on the multilayer representation of the network is useful to detect the timing of the changes during the time interval considered.
In our method, the temporal dynamic is captured in the two covariance matrices (eqs. \ref{eq:autocov_forw} \& \ref{eq:autocov_backw}) in terms of probabilities of following a given path and the RW rate plays the role of a filtering parameter that controls which spatio-temporal scales are considered.
However, two partitions cannot represent the entire dynamics in a time interval when the dynamics change multiple times. In this case, the interval can be sliced in several time windows and the flow stability applied on each slice. We show such an example in the next section.}

\subsubsection*{Free-ranging house mice contact network}
\label{sec:mice}

As a second example of real world application, we study an open population of house mice (\textit{Mus musculus domesticus}) living freely in a barn of approximately 72 m$^2$ near Zurich, Switzerland.
The barn is equipped with 40 nest boxes for the mice to rest and breed.  
Water and food are provided at twelve feeding trays inside the barn.
The activity of the mice is monitored thanks to subcutaneously implanted
radio-frequency identification (RFID) transponders and antennas situated
at the entrance of each nest box\cite{Konig2015}.
The time of the entering and leaving of the nest boxes are recorded along 
with the identity of the corresponding animal.
Male and female mice of at least 18\,g  are implanted with new transponders
with a unique RFID tag.
The presence of litters in the nest boxes is also monitored weekly.
The experiment has been initiated in 2002 and the continuous automatic 
reading and recording of the RFID transponders is in operation since 2007.
We use a dataset recording the mice activity from February 28 2017 to May 1st 2017
which capture the transition from winter to spring.
A temporal network is reconstructed with 437 nodes representing all the mice recorded
in the dataset and temporal events between two mice
representing their simultaneous presence in the same nest box.
There are more than 5.75 million events recorded with a millisecond resolution.
The distribution of event durations is very broad with a median at 64\,s, a 25 percentile at 7\,s and a 75 percentile at 6\,h\,25\,min.

\begin{figure}[ht]
\centering
 \includegraphics[width=\linewidth]{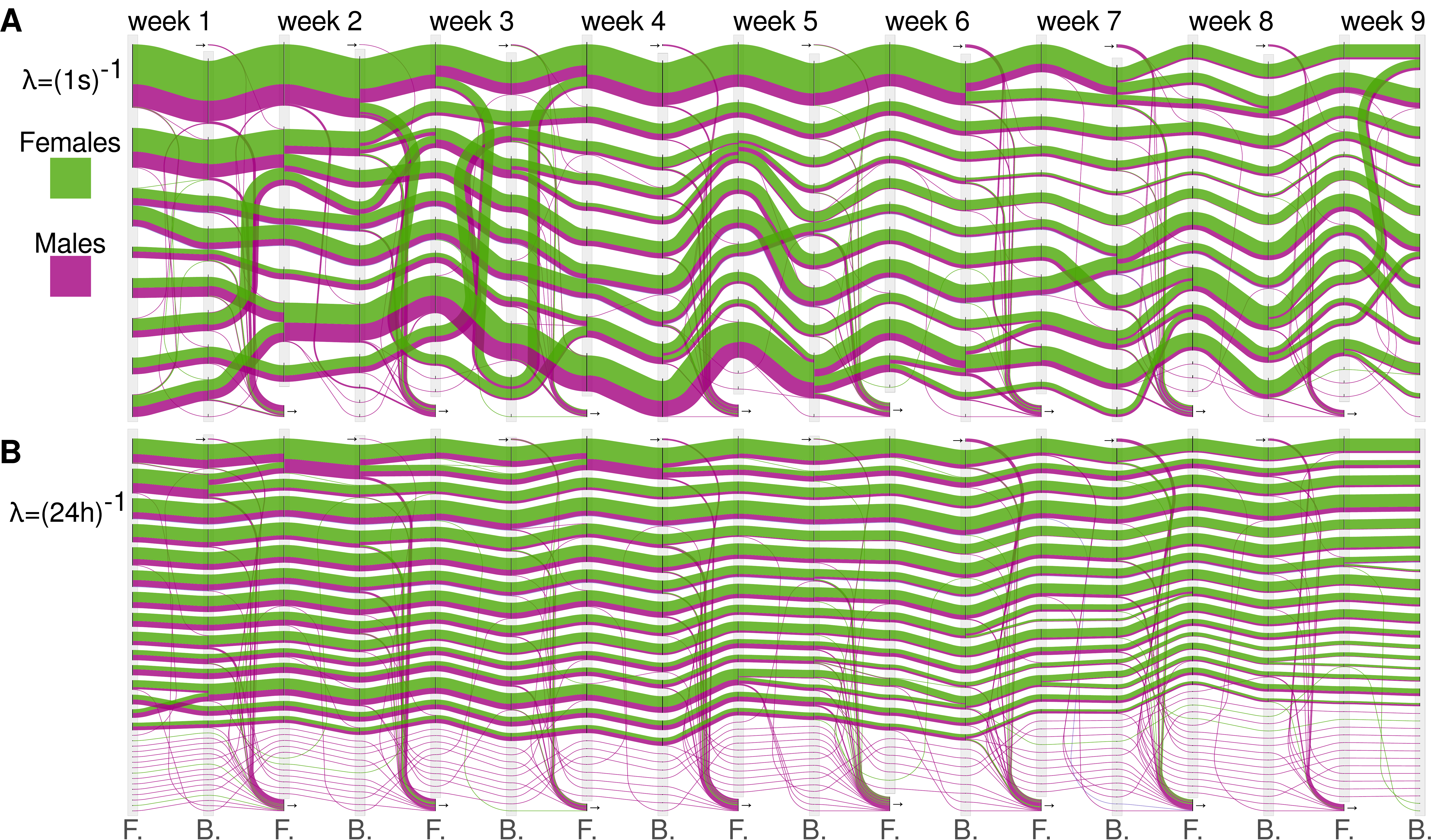}
 \caption{\textbf{Flow stability clustering of a contact network of free-ranging wild house mice.}
 (\textbf{A} \& \textbf{B}) Alluvial diagram representing the forward and backward partitions at two 
 different scales corresponding to random walk rates of  (1\,s)$^{-1}$ and  (24\,h)$^{-1}$ respectively.
 The flow corresponding to females is indicated in green and the one corresponding to males is in purple.
 The community dynamics at a rate of (1s)$^{-1}$ (\textbf{A}) reveals the existence of large communities
 during the first weeks corresponding to the end of February and beginning of March that split in smaller 
 communities as spring arrives.
 At a RW rate of (24h)$^{-1}$ (\textbf{B}) a finer description
of the dynamics is revealed with the presence of smaller social groups with compositions
and sizes that are very stable over the entire observation period.}
 \label{fig:mice_clust}
\end{figure}

\begin{figure}[ht]
\centering
 \includegraphics[width=0.9\linewidth]{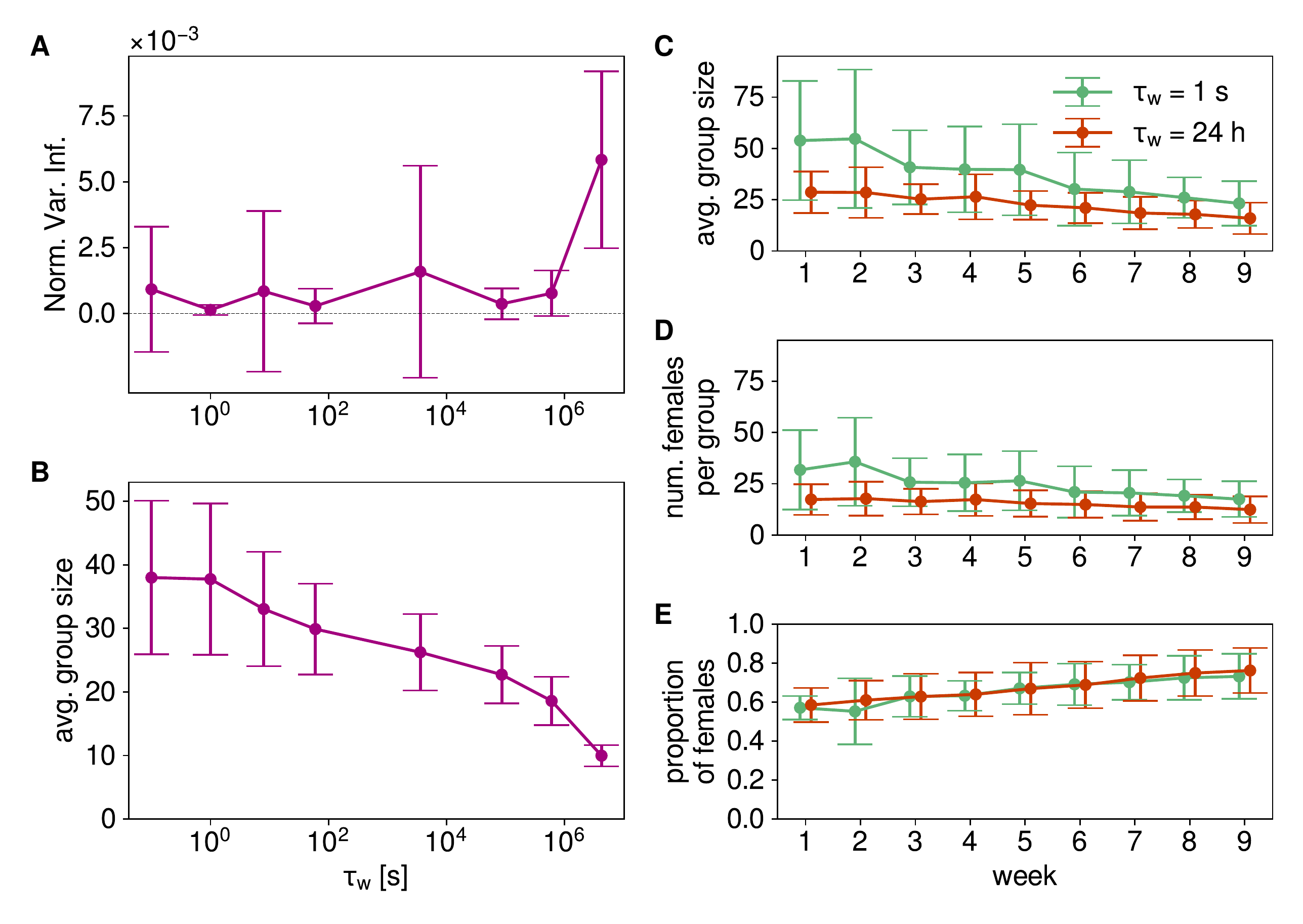}
 \caption{\textbf{Flow stability clustering statistics of the contact network of free-ranging wild house mice.}
 (\textbf{A}) Normalized Variation of Information 
 of the set of partitions  
 found by the stochastic optimization algorithm as a function of the random walk 
 characteristic waiting time.
 (\textbf{B}) Community sizes as a function of the random walk 
 characteristic waiting time.
 In (\textbf{A}) \& (\textbf{B}), the average and standard deviation computed across the 9
 pairs of forward and backward partitions are shown.
 (\textbf{C}) Group, or community, sizes as a function of the
 week. 
 (\textbf{D}) Number of females in each group as a function of the
 week. 
 (\textbf{E}) Proportion of females in each group as a function of the
 week.
 In (\textbf{C}), (\textbf{D}) \& (\textbf{E}), the average and standard deviation
across the forward and backward communities of each week are shown.}
 \label{fig:mice_nvi_stats}
\end{figure}

In order to observe the evolution of the community structure we divide
the period in 9 intervals of one week each.
For each week, we apply our method, using the linear approximation of the matrix exponential,
and vary the random walk rate to explore different dynamic scales.
This results in 9 pairs of forward and backward partitions that represent the evolution
in each week.
Figure \ref{fig:mice_clust} shows the nine backward and forward partitions represented
as an alluvial diagram for the random rates of $\lambda = (1\text{\,s})^{-1}$ (Fig. \ref{fig:mice_clust}A)
and $\lambda = (24\text{\,h})^{-1}$ (Fig. \ref{fig:mice_clust}B).
Figure \ref{fig:mice_nvi_stats}A \& B shows the Normalized Variation of Information (NVI) 
of the partitions and the number of groups (i.e. communities) found with 50 runs of the Louvain algorithm as a function of the
random walk characteristic waiting time ($\tau_w = 1/\lambda$).
The average and standard deviation taken over the 9 forward and backward partitions 
is displayed.
Minima in the NVI are visible for $\tau_w$ values of 1\,s, 60\,s and 24\,h.
These values indicate robust optimal partitions that correspond to intrinsic 
dynamic scales of the system.

The community dynamics at a rate of (1s)$^{-1}$ (Fig. \ref{fig:mice_clust}A) reveals the existence of large communities
with a high proportion of males (Fig. \ref{fig:mice_nvi_stats}E) during the first weeks corresponding to the 
end of February and beginning of March.
As spring arrives, the large groups split in smaller communities (see Fig. \ref{fig:mice_nvi_stats}C) and 
the proportion of females in groups increases as many males exit the system. 
In the mice population, the transition from winter to spring corresponds to a transition from low reproduction
to high reproduction\cite{Konig2012}.
In this case, there was no weaned pups sampled until April.
The average daily temperature in the barn also increased from freezing temperatures in February to 
temperatures around 20 $^\circ$C at the end of May.
The presence of larger groups in winter may be explained by the benefit of thermoregulation 
(winter huddles) and by the lower competition for reproduction\cite{Liechti2020}. 
At a RW rate of (24h)$^{-1}$ (Fig. \ref{fig:mice_clust}B) a finer description
of the dynamics is revealed with the presence of smaller social groups with compositions
and sizes that are very stable over the entire observation period (see Fig. \ref{fig:mice_nvi_stats}C).
While the average number of females per group stay extremely stable (see Fig. \ref{fig:mice_nvi_stats}D)
the proportion of males decreases similarly than for the coarser partition (Fig. \ref{fig:mice_nvi_stats}E)
suggesting that the females are forming the cores of the different social groups.

\rev{We compare these results with results obtained by two other dynamic community detection methods typically used in temporal network.
The first method consists of aggregating the activity over time windows to form a sequence of static networks. A static community detection method is then applied to each slice and the evolution of the communities from slice to slice is tracked.
Here, we use time windows of a half week, in order to have the same number of partition than with the flow stability method, and we follow the methodology of Liechti \textit{et al.}\cite{Liechti2020} that have studied the same mice population but over a different time frame. Communities are found at each slice with the hierarchical Infomap algorithm\cite{Rosvall2011}, and their evolution is tracked with an evolutionary clustering method\cite{Liechti2019}.
While this approach allows one to detect a coarse grained and fine grained evolution of the system (see Fig. S\ref{fig:mice_alluv_compa}), an issue arises as the method does not necessarily detect the same number of hierarchical level in all slices.
This renders the comparison of communities from slice to slice unclear. When tracking the number of communities per slice (see Fig. S\ref{fig:mice_alluv_compa}A) large variations are observed without knowing if they are due to real variations in the system or to the fact that the method found hierarchical levels at different scales.
The flow stability method, in addition to keeping temporal information within each time window, uses a resolution parameter with a physical meaning, the rate of the RW, which allows a principled comparison of slices at the same dynamical scale and results in a smooth variation of the number of communities per week (see Fig. \ref{fig:mice_alluv_compa}B).
The second method we compare our results with is the multilayer Infomap method applied to temporal networks\cite{DeDomenico2015,Aslak2018}. This approach allows one to perform a hierarchical clustering considering the entire network evolution and therefore find scales relevant across time points. 
We represent the contact network as a multilayer network with 18 layers being formed by the static aggregations with a window length of a half week. 
This approach detects 5 level of hierarchy, however the communities at each level are all elongated in time (see Fig. S\ref{fig:mice_multilay}) and the dynamics of splitting of the large communities in smaller communities is not recovered.
Here, we show that by using the flow stability method, we are able to retain temporal information within each slice and detect relevant dynamical scales revealing both the splitting of the communities in smaller groups at the arrival of spring and the existence of underlying smaller stable social groups.}

\subsubsection*{Uncovering the physical influences of network scientists}
\label{sec:aps}

As a last example, we demonstrate the possibility of our method to cluster non-stationary diffusion processes to investigate the diffusion of ideas in a network of co-authorship of articles published in journals of the American Physical Society (APS) between 1970 and 2010.
Scientists from many disciplines, including sociology, computer science and mathematics, contributed to the emergence of the academic discipline of network science. In the late 1990s-early 2000s, several physicists started to study complex networks and made a number of important contributions to the field.
We are interested in finding the influences in the field of physics that led these scientists to the study of complex networks.
The collaboration network has 194'451 nodes that correspond to authors and 
1'337'929 events corresponding to the co-authorship of two authors of the same article (see the Materials and Methods).
We consider that events represent collaborations between two authors
and set their length to 1 year and their ending times to the date of the article publication.
The event times are set on a monthly grid and we divide our investigation period
in decades.
We search all authors who have published an article in one of the APS journals between 2000 and 2010 with a keyword related to complex networks in the title or the abstract (the list of keywords is given in Supplementary Table S\ref{tab:complenet_keywords}).
We find 1'108 authors among which 1'048 are in the largest connected component of the network.
We compute a random walk process with an homogeneous initial condition on the 1'108 authors of complex networks articles starting in 2010. The initial probability distribution is zero on all other nodes.
We then let the random walk diffuse backward in time until 1970 with a characteristic waiting time of 10 years (eq. \ref{eq:temporal_trans_mat_rev}).
We compute the monthly inter-event transition matrices using the matrix exponential of the inter-event Laplacians.
For each decade, we find the best backward flow stability partition (eq. \ref{eq:back_int_stability}) using the probability distributions of the random walk process starting in 2010.
We assigned a main country based on most frequent country of their affiliations.
As APS journals Physical Reviews A,B,C,D and E are organized according to specific subjects in physics, we associate each author to the journal, among those 5, in which they published the most articles and use it as an indication of their main specialty in physics.
If an author only published in journals that cover the full scope of physics disciplines (e.g. Physical Review Letters and Review of Modern Physics), we associate them with a category "other".
When considering the backward diffusion process, over an interval $[t_1, t_2]$, $t_1<t_2$, the covariance matrix is non-null only for the nodes where $\vect{p}(t_2)>0$ (eq. \ref{eq:autocov_backw}). 
For each decade, we only consider authors who were active, i.e. who published at least one article during the decade, and who have a probability density at the end of the decade, i.e. $t_2$, superior than zero.

\begin{figure}[ht]
\centering
 \includegraphics[width=0.75\linewidth]{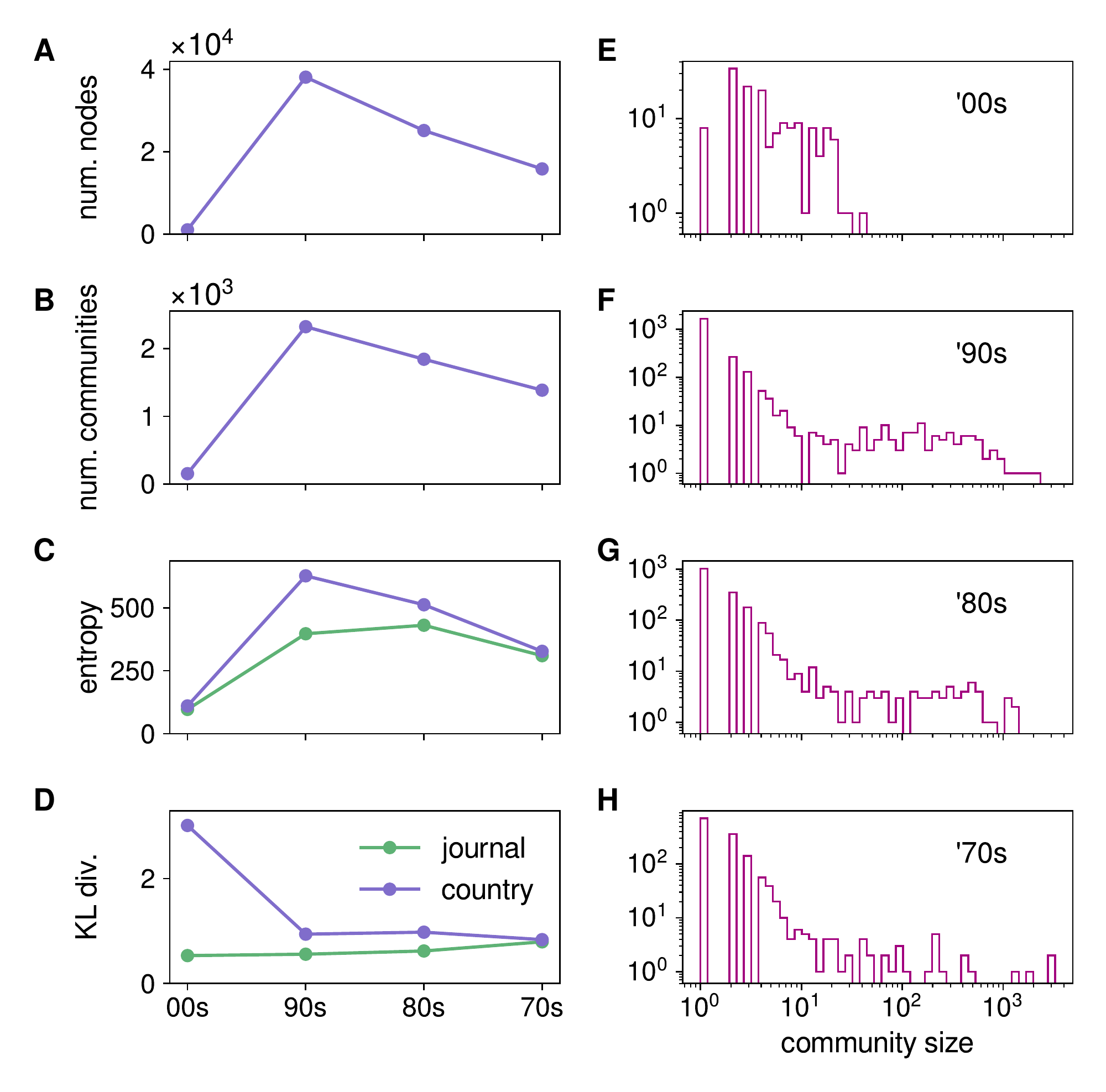}
 \caption{\textbf{Flow stability clustering statistics of the APS collaboration dataset.}
 (\textbf{A}) Number authors that were active and have a non-zero probability of participating in the diffusion process per decade.
 (\textbf{B}) Number of backward communities per decade.
 (\textbf{C}) Total entropy of the clustering.
 (\textbf{D}) Avergage Kullback-Leibler divergence of the clustering compared to the label distribution of the active nodes per decade.
 (\textbf{E}, \textbf{F}, \textbf{G} \& \textbf{H}) Histograms of the communities' sizes for each decade.}
 \label{fig:complenet_stats}
\end{figure}

Figure \ref{fig:complenet_stats} A \& B shows the number of nodes and communities for each decade revealing a drastic increase in 1990s compared to the initial condition in the 2000s. 
Although the process is diffusive and the support of the probability distribution expands as it evolves, \rev{because} the network size is decreasing as we goes back in time, the number of nodes considered decreases in the 80s and 70s.
Histograms of the community sizes for each decade are shown in Fig \ref{fig:complenet_stats} E to H.
We compute the total entropy of the clusterings with respect to the journal and country labels of each node which reveals that 
the diversity of the communities peaked in the 1990s and that
communities are in general more diverse in terms of country than journals (Fig. \ref{fig:complenet_stats}C).
This may be expected since edges in this network link authors publishing in the same journal.
To better understand how the diversity of the communities differ from each other, we compute the Kullback-Leibler divergence (KLD) of the clustering as the weighted average of the KLD between the distribution of labels of each community and the distribution of labels of the union of all communities, per decade (Fig. \ref{fig:complenet_stats}D).
The average KLD reveals that the distribution of countries in the initial communities from the 2000s is very different than the global distribution of countries of authors of complex networks articles. 
On the other hand, the average KLD of the distribution of journals is much smaller. 
Indeed, most of the initial authors (71\%) are associated to the journal Physical Review E and therefore the communities do not show a large diversity in terms of journals, however the large KLD reveals that they are very diverse in terms of country distribution (the most common author country is the USA, with 20\% of the authors).
As the diffusion process moves backward in time, the average KLD of the country distributions stays larger than the KLD of the journal distributions, however their difference becomes smaller and smaller.

\begin{figure}[ht]
\centering
 \includegraphics[width=\linewidth]{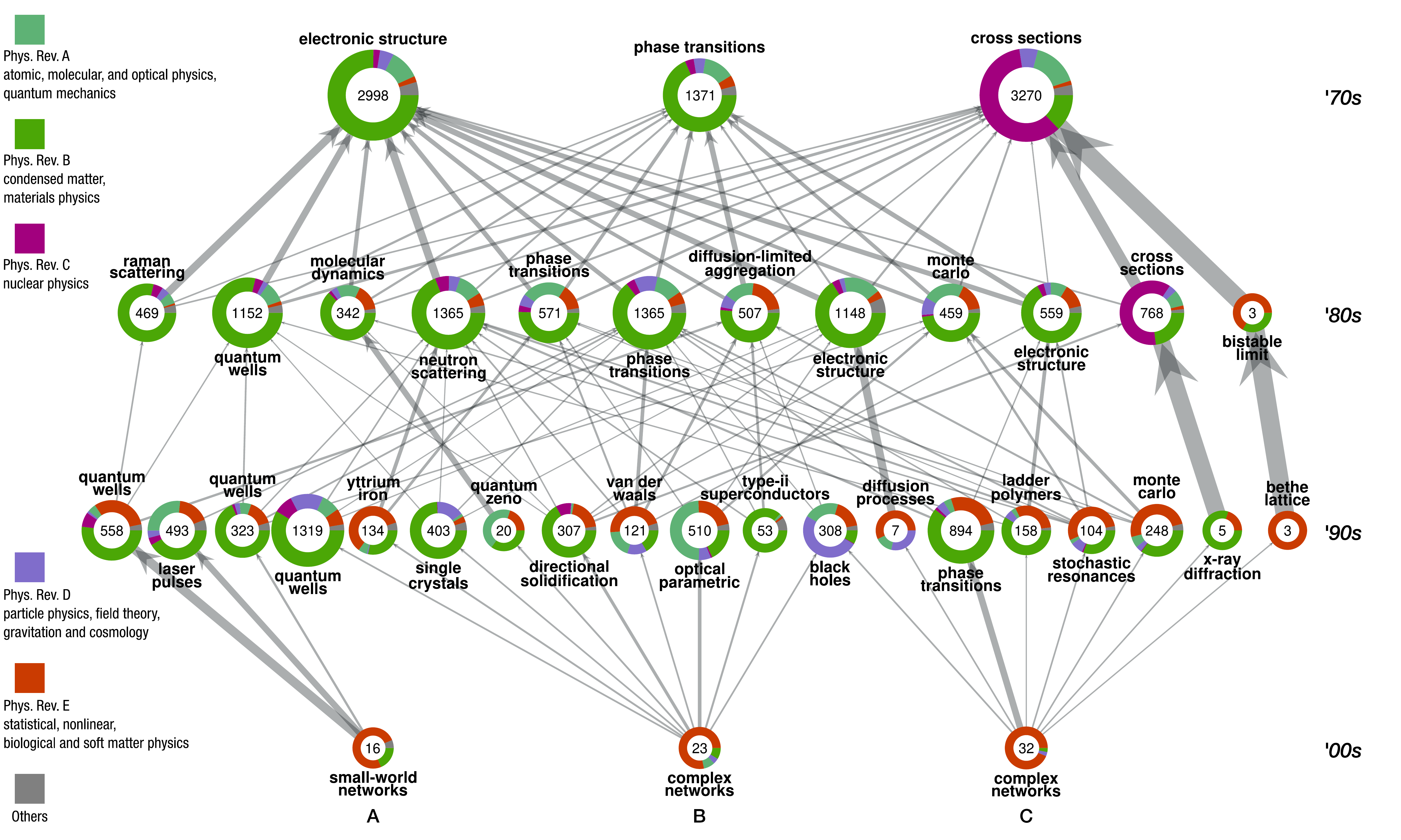}
 \caption{\textbf{Influential communities of authors of articles published in the APS journals for three communities of network scientists in the 2000s.}
 Each node represent a community and its size is indicated in the center.
 The colors represent the distribution of journals inside each community where each author is associated to the journal in which they published the most (excepted large scope journals).
 The pair of words next to each node indicate one of the most frequent pair of words of all the titles of the articles belonging to the community.
 Arrows between the communities represents probability transitions  ($>5\%$) from community to community of the diffusive process starting in 2010 and finishing in 1970.}
 \label{fig:complenet_flow}
\end{figure}

It is interesting to understand the relation between communities of different decades.
Here, contrary to the previous examples, we are not interested in necessarily following the same nodes across intervals to understand how communities evolved, but rather in following the diffusion process. 
We can link communities from one decade to another by clustering the transition matrix computed between those two decades (see the Materials and Methods section) and following the transitions with the highest probabilities.
To illustrate this process, we selected three initial communities from the 2000s that had different country distributions: community A (50\% USA, 44\% Hungary), community B (30\% UK, 22\%Finland, 17\% Spain) and community C (34\% Italy, 22\% USA, 19\% Spain). The names, countries and journals of all authors in these communities is given in Supplementary Table S\ref{tab:aps_init_authors}.
The list of "ancestor" communities of the 1990s is found as the communities towards which the transition probability of the random walk process, starting from one of the three initial communities, is larger than 5\%. Similarly, we find the "ancestor" communities of the 1980s and 1970s.
Figure \ref{fig:complenet_flow} shows the three initial communities and their "ancestor" communities at each decade along with the transition probabilities between each community and the distribution of authors' main journals in each community.
Supplementary Figure S\ref{fig:complenet_flow_countries} shows the communities together with the distribution of authors' main countries in each community.
We discover that the three initial communities have different influence communities in the '90s (only communities B and C have one common ancestor at this stage) when considering transition probabilities larger than 5\%.
Community A has only three ancestors in the '90s that are dominated by Phys. Rev. B (condensed matter and material physics) but with different distributions of secondary journals.
The most frequent pair of words in the articles' titles reveal that two communities are mostly focused about \emph{quantum wells} and the third one about \emph{laser pulses}.
Figure S\ref{fig:complenet_flow_countries} shows that the two quantum wells communities differ in their country distribution, one of them having a large portion of authors with affiliations in the UK.
Community B has the largest number of ancestors in the '90s that are also the most diverse in terms of journal distributions. The main topics of each communities are also very diverse, ranging from \emph{Van der Waals} forces to \emph{black holes}.
Finally, community C has also a wide range of influences in the 90s that is dominated by the journals Phys. Rev. B and Phys. Rev. E (statistical, nonlinear, biological and soft matter physics) with topics such as \emph{diffusion processes}, \emph{phase transitions} and \emph{Monte Carlo} methods.
As we follow influences in the 80s and 70s, more common ancestor communities are found that have all a significant proportion of Phys. Rev. B but focused on different topics.
Three communities are found in the 1970s: two have relatively similar journal compositions (dominated by Phys. Rev. B) but focus on different topics (\emph{electronic structure} and \emph{phase transitions}) and the third one is dominated by Phys. Rev. C (\emph{nuclear physics}) and the topic of \emph{cross sections}.
The third community is only a significant ancestor (probability of transition $>5\%$) of community C.
We note that the three communities of the '70s are among the four largest communities of this decades (see Fig. \ref{fig:complenet_stats}H). 
This can be expected as larger communities have a higher probability of being on a random walk, but this also reveal that there is another community, as important as those three, that is not a significant influence for these three initial communities of network scientists.
Note also that in order to study the transmission of influences from the 1970s to the 2000s using the same diffusion process, one would compute the inverse forward communities based on $\matr{S}_\textrm{forw}^\textrm{inv}$.
With this example, we demonstrate an original usage of our method for clustering non-stationary processes in temporal networks that allows us to uncover new insights about the influences, in the field of physics, of network scientists.

\section*{Discussion}

The classical static definition of communities as clusters of densely connected nodes
does not generalize well to temporal networks without resorting to temporal aggregations over some time-windows
to evaluate the ``connectedness'' of groups of nodes.
In many cases, this aggregation does not prevent the detection of 
communities and their temporal evolution.
However, many processes can be occurring simultaneously in a system
described by a temporal network and each of them at different rates.
We showed that the aggregation of temporal networks 
over time windows can lead to a loss of information at certain dynamical scales
and render the detection of processes occurring at certain scales impossible. 
Here, we propose a framework based on the clustering of the flow of 
random walkers evolving with the network that allows us
to define communities in temporal networks while keeping 
temporal information of time-respecting paths, without resorting to temporal
aggregation and without assuming the existence of a stationary state \rev{of the flow}.
To capture the asymmetric relations between nodes due 
to the temporal evolution of the network, we describe the communities 
over a given time interval with two partitions:
the \emph{forward} partition that groups nodes in the same community
if the flow of random walkers starting on them tend to stay together
until the end of the interval
and the \emph{backward} partition that groups nodes in the same community
if the flow of random walkers that ends on them tended to stay together since 
the beginning of the interval.
Time symmetry is an essential concept in theoretical physics, 
associated to energy conservation through Noether’s theorem and 
to the emergence of an arrow of time through thermodynamics. 
While this work does not aim at modeling a physical system directly, 
it provides an interesting viewpoint that should be explored further.
Indeed, we model systems that show time asymmetry at the microscopic level,
the random walk process being diffusive and non-reversible in general, 
yet that can capture time symmetry at the mesoscopic level of communities when the 
forward and backward partitions are similar.

Our framework provides a natural way to explore the different natural dynamical
scales present in a system by varying the rate of the random walks which
plays the role of a dynamical resolution parameter.
In terms of the classification by Rossetti and Cazabet\cite{Rossetti2018},
each partition taken alone could be classified in the temporal trade-off
category. The forward partition depends on the network topology
at ``time $t$'' and also on the future topology, while the backward
partition depends on the topology at ``time $t$'' and in the past.
The two partitions taken together could then be classified in the cross-time
category, depending on the entire evolution of the network in a
given time-interval.
The temporal flow stability is also a natural generalization of static networks 
concepts such as modularity and Markov Stability and draws links with 
clustering methods for directed networks (see Supplementary Text  \nameref{sec:co-clust}).
An advantage of our method is that, for a given time interval, the method has only one parameter
with a principled meaning, the random walk rate,
while other approaches may require the tuning of several parameters
(e.g. slice resolution parameter and inter-slice coupling\cite{Mucha2010}).
In static networks, the concept of Markov stability has already been expressed
in terms of a filtering process in the framework of graph signal processing\cite{Tremblay2014,Gutierrez2020}.
Here, the random walk process can be seen as a spatio-temporal filter on the temporal network that 
weights the importance of interactions depending on their
duration and frequency.
Other types of filters could be design to focus on particular
processes such as cyclical activity for example.
\rev{The usage of different Laplacians, defining different diffusion processes, or time kernels modulating the importance of different temporal patterns in the objective functions could be used to design new methods.}
Our framework opens the door for the definition of new concepts for temporal networks
in terms of RW probabilities and flows that may help to disentangle 
the complex processes simultaneously occurring in systems described as
temporal networks.

\section*{Materials and Methods}

\subsection*{\rev{Flow modeling}}

We consider the temporal network with $N$ vertices and $M$ undirected events
defined in section \nameref{sec:temp_net}.
We define the ordered set of distinct event times, $T^i$, as the union
of the sets of starting times, $T^s$, and ending times, $T^e$.
The event times effectively defines new events at a higher temporal 
resolution such that there is no change in the network between two consecutive 
times \rev{(e.g. in Fig. \ref{fig:schema}A the event times are indicated by black dots)}.
One can compute the transition matrix 
between two arbitrary times from the product of the transition matrices 
for each inter-event time interval.
On this new temporal grid, one finds for the transition probability matrix 
between to arbitrary times $t_1$ and $t_2$ ($t_1 < t_2$)
\begin{equation}
\matr{T}(t_1,t_2) = \matr{\hat{T}}(t_1, t_m) \left[\prod_{k=m}^{n-1}\matr{\hat{T}}(t_k, t_{k+1})\right]\matr{\hat{T}}(t_n, t_2), 
\label{eq:temporal_trans_mat}
\end{equation}
with $m<n$, $t_m\geq t_1$ being the time of the first event after, or at, $t_1$ and $t_n<t_2$ the time 
of the last event before $t_2$.
To compute the transition matrix corresponding to the time-reversed evolution of the network, from $t_2$ to $t_1$, 
we perform the matrix product in the reversed order:
\begin{equation}
\matr{T}_\textrm{rev}(t_2,t_1) = \matr{\hat{T}}(t_2, t_n) \left[\prod^{m+1}_{k=n}\matr{\hat{T}}(t_{k}, t_{k-1})\right]\matr{\hat{T}}(t_m, t_1). 
\label{eq:temporal_trans_mat_rev}
\end{equation}

\label{rev:chapman_kol}
In order to ensure that the transition probability matrix satisfies the Chapman-Kolmogorov equation
$\matr{T}(t_1,t_3) = \matr{T}(t_1,t_2)\matr{T}(t_2,t_3)$ for arbitrary times $t_1<t_2<t_3$,
one must ensure that in particular $\matr{\hat{T}}(t_k,t_{k+1}) = \matr{\hat{T}}(t_k,t_l)\matr{\hat{T}}(t_l,t_{k+1})$
where $t_k < t_l < t_{k+1}$ and $t_k$ and $t_{k+1}$ are consecutive times on the high resolution temporal grid.
Assuming that walkers have a constant probability of jumping per unit of time given by the rate $\lambda$,
this is uniquely satisfied by the solution $
 \matr{\hat{T}}(t_k, t_{k+1}) = e^{-\lambda\matr{L}(t_k)\tau_k},
$
with $\tau_k = t_{k+1} - t_k$ and where
$
\matr{L} = \matr{I} - \matr{D}(t)^{-1}\left(\matr{A}(t)+\matr{S}(t)\right)
$
is the \emph{random walk graph Laplacian} at time $t$, $\matr{A}(t)$ is the adjacency
matrix at time $t$, $\matr{S}(t)$ is the self loops matrix at time $t$, with zeroes
everywhere except on the diagonal element $i$ corresponding to nodes with zero out-degree, $k(t)_i^{}$,
$\matr{D}(t)$ is the diagonal matrix with $\matr{D}(t)_{ii} = k(t)_i^{}$ 
if $k(t)_i^{}>0$ and $\matr{D}(t)_{ii} = 1$ otherwise.
The element $(i,j)$ of $\matr{L}(t)$ is therefore given by
\begin{equation}
(\matr{L}(t))_{ij} =
\begin{cases}
-\frac{a(t)_{ij}}{\max(k(t)_i^{},1)} & \text{if } i\neq j,\\
1 - \delta(k(t)_i^{}, 0)  & \text{if } i=j,
  \end{cases}.
\end{equation}

We have $\matr{L(t)}\vect{1}=0$, i.e $\vect{u_1} = \frac{1}{N}\vect{1}$
is a right-eigenvector of $\matr{L}$ associated with the eigenvalue $\epsilon_1 = 0$.

Note that for $t>0$, $e^{-\lambda\matr{L}(t_k)\tau_k}$ may contain 
non-zero non-diagonal terms that are equal to zero in $\matr{L}$ 
(or $\matr{A}$), i.e. $e^{-\lambda\matr{L}(t_k)\tau_k}$ takes into 
account trajectories with multiple steps.


\subsection*{Covariance of non-stationary random walks}

To find a relevant partition of the nodes between two time points $t_1$ and $t_2$ ($t_1<t_2$),
we consider the covariance of a flow of random walkers, 
performing a Continuous Time Random Walk (CTRW)\cite{Montroll1965}
on the network constrained by the activation of edges,
between the different clusters\cite{Delvenne2010,Lambiotte2014}.
A partition that is well aligned with this flow will correspond to high values of the covariance inside each cluster.

Following the framework of the stability of a network partition\cite{Delvenne2010} but in the case of a temporal network 
and without assuming an ergodic and reversible Markov chain with a stationarity distribution,
we assign a different real value $\alpha_i$ ($i=1,...,c$) to the vertices of each of the $c$ clusters
and consider the values $\alpha_i$ observed by a random walker as a stochastic process
$(X_t)_{t\in\mathbb{R}}$ which is not necessarily Markovian and not necessarily stationary.
The covariance of this process evaluated between $t_1$ and $t_2$ is given by
\begin{equation}
\mathrm{cov}\left[X(t_1)X(t_2)\right] = \mathrm{E}\left[X(t_1)X(t_2)\right] - 
\mathrm{E}\left[X(t_1)\right]\mathrm{E}\left[X(t_2)\right],
\end{equation}
where $\mathrm{E}\left[X(t)\right]$ represents the expectation of the random variable $X(t)$. 

Introducing $\vect{p}(t)$, the $1\times N$ row-vector with element $p_i(t)$ equal to 
the probability of finding a random walker on node $i$ 
at time $t$, and \rev{using} the $N\times N$ transition matrix $\matr{T}(t_1, t_2)$ \rev{defined in eq. (\ref{eq:trans_mat})} where 
element $(i,j)$ is equal to the conditional probability for a random walker 
to be on node $j$ at $t_2$ if it was on node $i$ at $t_1$, we find
\begin{align}
\mathrm{cov}\left[X(t_1)X(t_2)\right] & =  \sum_{i=1}^N\sum_{j=1}^N \alpha_i p_i(t_1) T_{i,j}(t_1,t_2) \alpha_j -
\left(\sum_{i=1}^N \alpha_i p_i(t_1) \right)\left(\sum_{i=1}^M \alpha_i p_i(t_2)\right)\notag\\
& =\sum_{\ell,m=1}^c \alpha_\ell \alpha_m \sum_{i,j=1}^N \delta(\ell,c_i)\left( p_i(t_1) T_{i,j}(t_1,t_2) -
p_i(t_1) p_j(t_2)\right)\delta(c_j,m)\notag\\
& = \vectg{\upalpha}^\mathsf{T} \matr{R}(t_1, t_2) \vectg{\upalpha},
\end{align}
where $\vectg{\alpha}$ is the $1\times c$ column vector of labels of the $c$ communities and
\begin{equation}
\matr{R}(t_1, t_2; H) = \matr{H}^\mathsf{T}\left[\matr{P}(t_1) \matr{T}(t_1, t_2) - \vect{p}(t_1)^\mathsf{T} \vect{p}(t_2)\right]\matr{H}
= \matr{H}^\mathsf{T}\matr{S}(t_1, t_2)\matr{H}
\label{eq:clustered_autocov} 
\end{equation}
is the $c \times c$ clustered covariance matrix between $t_1$ and $t_2$ with $\matr{P}(t) = \mathrm{diag}(\vect{p}(t)) \forall t \in [t_1,t_2]$
and 
\begin{equation}
    \matr{S}(t_1, t_2) = \matr{P}(t_1) \matr{T}(t_1, t_2) - \vect{p}(t_1)^\mathsf{T} \vect{p}(t_2)
\end{equation}
is the $N \times N$ covariance matrix between $t_1$ and $t_2$.
Importantly, $\matr{R}(t_1,t_2;H)$ only depends on the network and its partition, and not on the specific, yet arbitrary, values of $\vectg{\upalpha}$.

This expression can then be used to find a partition clustering the covariance in blocks
where the random walkers are likely to remain for a long time, i.e. where the covariance 
is high.
In the case of static networks, this expression reduces to 
the framework of Markov stability\cite{Delvenne2010,Lambiotte2014}, 
where the random walkers eventually reach a stationary distribution.
See Supplementary Texts: \nameref{sec:rel_static_networks}, \nameref{sec:co-clust} \revv{\& \nameref{sec:special_cases} as well as Supplementary Table S\ref{tab:relations}} for
the relations between our approach and well-known static networks heuristics such as modularity optimisation.
In the case of temporal networks, the activity-driven model has been used to
approximate a stationary distribution and generalize the Markov stability framework\cite{Petri2014}.
Another approach for temporal networks consists of treating them as 
multilayer networks and considering a random walk that moves insides
layers as well as in-between layers effectively disregarding the direction
of time and the causality of random walkers' paths\cite{Mucha2010}.

\subsection*{Linearisation of the transition matrix \rev{and computation of the covariances}}

As the computation of the matrix exponential can be relatively time costly for large network, we
introduce a linearisation of eq. (\ref{eq:trans_mat}) using two linear interpolations
\begin{equation}
\tilde{T}(t_1,t_2;\tau_w) =
\begin{cases}
 (1-\lambda^\star) \matr{I} + \lambda^\star\matr{T}_{\textrm{DT}} & \text{ for } 0\leq \lambda^\star \leq 1, \\
 \frac{1}{1-\lambda^s}[(\lambda^\star-\lambda^s)\matr{T}_{\textrm{DT}} + (1-\lambda^\star)\matr{W}] & \text{ for } 1 < \lambda^\star  \leq \lambda^s, \\
 \matr{W} & \text{ for } \lambda^\star > \lambda^s,
 \end{cases}
\label{eq:lin_trans}
\end{equation}
where $\matr{T}_{\textrm{DT}}=\matr{I}-\matr{L}$ is the one-step discrete time random walk transition matrix, $\lambda^\star =\frac{t_2-t_1}{\tau_w}$,
$\matr{W} = \lim_{n\rightarrow\infty} \matr{T}_{\textrm{DT}}^n$  is the limiting transition matrix
and $\lambda^s=t_s/\tau_w$, with $t_s$ the time taken by the random walk to reach stationarity. In all the examples in this article, we use $\lambda^s=10$.

In order to compute the linear approximation of the transition matrix $\matr{T}(t_1,t_2)$
of a time-evolving network (eq. \ref{eq:temporal_trans_mat}),
we first compute the linear approximation of each inter-event transition 
matrix with eq. (\ref{eq:lin_trans}).
The limiting transition matrix $\matr{W}$ can be easily computed for 
undirected network. The matrix $\matr{W}$ has non-zero values only
in diagonal blocks that correspond to each connected component of the graph.
The stationary distribution of the $n^\text{th}$ connected component
is $\vect{\pi}_n$, with element ${\vect{\pi}_n}_i=k_i/N_n$,
where $k_i$ is the degree of node $i$ and $N_n$ is the size of component $n$.
The $N_n$ rows of the $n^\text{th}$ block of $\matr{W}$ are then all copies of 
the vector $\vect{\pi}_n$.

\label{rev:complimit}
\rev{For large networks, our method is limited to cases where the number of edges being active simultaneously remains small, which is usually the case in temporal networks.
In this case, we find that the computations are greatly simplified by the fact that inter-event Laplacians are usually extremely sparse and one can compute the matrix exponential of each connected component independently.}

\sloppy
\rev{The integral of the forward covariance is obtained by performing the integral $\int_{t_1}^{t_2}(\Tmat(t,t_1)\Pmat(t)^{-1/2})(\Tmat(t,t_1)\Pmat(t)^{-1/2})^\mathsf{T}dt$ and then multiplying its rows and columns by $\pvec(t_1)$.
The integrand can be efficiently computed as a sparse gram matrix and only its upper (or lower) triangular values needs to be computed and stored as it is symmetric,
The outer product of $\pvec(t_1)$ in eq. (\ref{eq:autocov_forw}) is a rank 1 matrix and therefore can be efficiently stored using only a vector.}

\rev{The computational cost is small for network sizes $N$ where $N\times N$ matrices can be stored in memory (e.g. $\sim$ 6\,GB for a $N=4\times10^4$ double precision floats symmetric matrix). 
For large networks, the main limitation is the fact that the total transition matrix and the integral of the covariance may start to become less sparse. All elements of these matrices that are inside connected components have a non-zero value. As the integration interval increases, for very large networks, the connected components sizes increases which slows down the matrix operations and may require large memory storage.
This is a limit of our method and to scale it to larger networks, we keep the matrices sparse by neglecting RW paths with very low probabilities.
We keep only values of these matrices with probabilities above a certain value. We applied this strategy in the case of the physical influences of network scientists (see below).}

\subsection*{Tracking scientific influences in the American Physical Society co-authorship dataset}

Similarly to Ref. \cite{Sinatra2016}, we consider only articles having 10 or less authors in the APS dataset in order to exclude articles from "big science" projects that do not correspond to the concept of collaboration that we are investigating. We also consider only articles with at least two authors since we are interested in the diffusion of ideas between co-authors.
We use the author name disambiguation provided in Ref. \cite{Sinatra2016}.
The countries corresponding to the authors' affiliations were extracted by first examining the affiliations' most common trigrams and bigrams that contain names of known institutions and locations.
This allows us to extract the countries corresponding to 96\% of the affiliations.
The countries of the remaining affiliations are extracted by using three approaches: with the a named entity extraction library (\url{https://github.com/iwpnd/flashgeotext}),
by fuzzy matching the bigrams and trigrams (allowing the n-gram Jaro–Winkler similarity to be $\geq 0.95$) to allow slight mispellings and finally by using the OpenStreetMap Nominatim geocoder (\url{https://github.com/geopy/geopy}).
Over the initial 224'992 unique affiliations, we were unable to assign a country to only 89 affiliations.
The full mapping and the code used to produce it are available at \url{https://doi.org/10.7910/DVN/I87AXV}.

We compute the inter-event transition matrices without the linear approximation on a monthly resolution. We use sparse matrix representations and compute the matrix exponential on each connected component of the Laplacian matrix in parallel in order to limit memory and computation time.
Moreover, we threshold the transition matrices (values smaller than $1\times10^{-6}$ of the maximum) and of covariance integrals (absolute values smaller than $1\times10^{-9}$ of the maximum) to further limit memory usage.

The transition probabilities between the backward communities of 
decade $d$ and the ones of the previous decade $d-1$ are computed as 
\begin{equation}
    \matr{T}_{d\rightarrow d-1} = \matr{C}_{b,d}^{-1} \matr{H}_{b,d}^\mathsf{T} \matr{T}(t_d, t_{d-1}) \matr{H}_{b,d-1},
\end{equation}
where $\matr{H}_{b,d}$ $d\in$ \{'00s, '90s, '80s, '70s\} are the indicator matrices encoding the backward communities and $\matr{T}(t_d, t_{d-1})$ is the transition matrix of the random walk process starting at the end of the decade $d$ and ending at the end of decade $d-1$, e.g. 2010 to 2000 for $d=$'00s and $d-1=$'90s. 
The matrix $\matr{C}_{b,d}^{-1}=\textrm{diag}\left({\matr{1}\matr{H}_{b,d}}\right)$ is the diagonal matrix containing the sizes of each communities in $\matr{H}_{b,d}$. Finally, the transition probabilities between the '00s and all earlier decades are found by multiplying, in time reversed order, the matrices for each decade.

\noindent \textbf{Acknowledgements:} 
The authors thanks Barbara K\"{o}nig and Anna Lindholm at the University of Zurich for the access to the wild mice dataset and the helpful and interesting discussions.
The authors thank Benjamin Chiêm, Matteo Cinelli, Mauro Faccin, Leonardo Gutierrez, Jonas I. Liechti, Alexey Medvedev, Leto Peel and Michael Schaub for the fruitful discussions.
\noindent \textbf{Funding:} A. B. thanks the Swiss National Science Foundation for the financial support (Grant P300P2\_177793).\\
\noindent \textbf{Author Contributions:} All authors conceived the project. A.B. developed the theoretical framework with support from all authors, performed the simulations, analysis, implemented the computer code and wrote the manuscript. All authors reviewed and contributed to the final manuscript.\\
\noindent \textbf{Competing Interests:} The authors declare that they have no competing interests.\\
\noindent \textbf{Data and materials availability:}
The primary school contact network is available from SocioPatterns at \url{http://www.sociopatterns.org/datasets/}.
The wild mice contact network is available at \url{http://doi.org/10.5281/zenodo.4725155}.
The APS dataset can be requested at \url{https://journals.aps.org/datasets}.
The author name disambiguation of the APS dataset is available in the Supplementary Material of Ref. \cite{Sinatra2016} at \url{https://doi.org/10.1126/science.aaf5239}.
\rev{Code and additional data allowing to replicate all the results in this article are deposited in the Harvard Dataverse repository at \url{https://doi.org/10.7910/DVN/I87AXV}}.
A python code implementing the flow stability framework is available at \url{https://github.com/alexbovet/flow_stability} \rev{and deposited in the Zenodo repository with DOI \texttt{10.5281/zenodo.5786949}}.

\clearpage

\renewcommand{\theequation}{S\arabic{equation}}

\captionsetup[figure]{labelformat=sfiglab}
\captionsetup[table]{labelformat=stablab}

\setcounter{figure}{0} 
\setcounter{equation}{0} 
\setcounter{table}{0} 

\begin{center}

{\Large Supplementary Materials for\\[0.5cm]
\textbf{Flow stability for dynamic community detection}}\\[0.5cm]
 
 {Alexandre Bovet$^{\star}$, Jean-Charles Delvenne, Renaud Lambiotte}
 
 {\footnotesize $^{\star}$Corresponding author. Email: alexandre.bovet@maths.ox.ac.uk}
 
\end{center}

\section*{Supplementary Text}

\subsection*{Relations with community detection in static networks}
\label{sec:rel_static_networks}

The expression of the clustered-covariance eq. (\ref{eq:clustered_autocov}) encompasses several well-known heuristics for the
clustering of static network as special cases.
The simplest example is the case of an undirected static network with $M$ edges described by the adjacency 
matrix $\matr{A}$ (see Supplementary Text \nameref{sec:co-clust} and Tab. S\ref{tab:relations} for more examples).
Considering a discrete time random walk on this network,
the transition matrix after $n$ steps is given by $\matr{T}(n)=(\matr{D}^{-1}\matr{A})^n$ where $\matr{D}$ 
is the diagonal matrix with diagonal element $(i,i)$ equal the the degree $k_i$ of vertex $i$.
The stationary distribution of the random walk is given by the vector $\vectg{\pi}$, with elements $\pi_i=k_i/2M$.
In this case, the element $(i,j)$ of the clustered covariance 
(eq. \ref{eq:clustered_autocov}), computed after one step and evaluated at stationarity, reduces to
\begin{equation}
R_{ij}(n=1; H) = \left(\pi_i \frac{A_{ij}}{k_i}-\pi_{i}\pi_{j}\right) \delta\left(c_{i}, c_{j}\right)
= \frac{1}{2 m}\left(A_{i j}-\frac{k_{i} k_{j}}{2 m}\right) \delta\left(c_{i}, c_{j}\right).
\end{equation}

Or in matrix notation
\begin{equation}
\matr{R}(n=1; H) = \matr{H}^\mathsf{T}\left[\matr{\Pi} \matr{D}^{-1}\matr{A} - \vectg{\uppi}^\mathsf{T} \vectg{\uppi}\right]\matr{H} = 
\frac{1}{2 m}\matr{H}^\mathsf{T} \matr{B}\matr{H},
\label{eq:static_modul}
\end{equation}
where $\matr{B}$ is the modularity matrix\cite{Newman2006,Newman2006b}.
We recognize the classical Newman-Girvan modularity\cite{Newman2004} by taking the trace of the 
clustered covariance: $Q = \textrm{trace}\left[\matr{R}(n=1; H)\right]$\cite{Delvenne2010}.
Finding a partition that maximizes the modularity, i.e. the number of observed edges inside 
each clusters minus the number expected from a random null model, can then be seen as finding a partition that 
the maximizes the elements on the diagonal of $\matr{R}(n=1; H)$, namely the probability that
the walkers stay in the same clusters after one step minus the same probability for two independent
walkers, evaluated at stationarity.
This analogy allows us to see that the random null model of the modularity corresponds to the outer product
of the stationary distribution of the random walks.
As a matter of fact, the stationary distributions of different models of random walks correspond 
to different generative network null models\cite{Lambiotte2014}.
This random walk framework has been shown to be a very fruitful way to generalize modularity 
optimization and unify different clustering heuristics\cite{Delvenne2010,Schaub2012,Lambiotte2014}.
By allowing the walkers to make multiple steps\cite{Delvenne2010},
or by considering a continuous time random walk\cite{Lambiotte2014}, one can use the 
elapsed time of the random walk as a resolution parameter allowing to recover the 
multiscale community structure of networks\cite{Schaub2012} and overcome the resolution limit of 
the Newman-Girvan modularity\cite{Fortunato2007}.

A particularly interesting special case of application of eq. (\ref{eq:clustered_autocov})
is the case of a static directed network 
with $M$ edges and with adjacency matrix $\matr{A}$.
The in-degrees of node $i$ is $k_{i}^{\textrm{in}}=\sum_j A_{ji}$ and its out-degree is 
$k_{i}^{\textrm{out}}=\sum_j A_{ij}$.
The transition matrix after one step is given by $T_{ij}=A_{ij}/k_{i}^{\textrm{out}}$ if 
$k_{i}^{\textrm{out}} \neq 0$ and  $T_{ij}=0$ if $k_{i}^{\textrm{out}} = 0$. 
Considering an initial distribution of walkers given by $p_i(0)=k_{i}^{\textrm{out}}/M$, 
the distribution after one step is given by $p_i(1)=\sum_j p_j(0)T_{ij}=k_{i}^{\textrm{in}}/M$.
Replacing these expressions in eq. (\ref{eq:clustered_autocov}) and taking the trace of the
clustered covariance matrix, we find
\begin{equation}
\textrm{trace}\left[\matr{R}(n=1; H)\right] = \frac{1}{m}\sum_{ij}\left(A_{ij} - 
\frac{k_{i}^{\textrm{out}} k_{j}^{\textrm{in}}}{m}\right)\delta\left(c_{i}, c_{j}\right) = Q^\textrm{d.},
\end{equation}
which is a classical generalization of modularity to directed networks\cite{Arenas2007,Fortunato2010}.

\subsection*{Relations with co-clustering}
\label{sec:co-clust}
It is interesting to note that the clustering of symmetric covariance matrices 
with the Markov stability framework is linked to the spectral approaches of graph clustering\cite{Delvenne2010}.
Indeed, as the time parameter increases, the contribution of the eigenvectors of the
transition matrix, which are similar to the ones of the random walk graph Laplacian, $\matr{L}$\cite{Masuda2016}, 
to the stability are re-weighted according to their eigenvalues to give more weight to
larger and larger scales in the network.
In the static undirected case, the random walk has a stationary distribution, $\vectg{\pi}$,
i.e. $\vectg{\pi}$ is a left-eigenvector of $\Tmat$ with eigenvalue 1.
The covariance is given by $\Smat_{static}(\tau)=\matr{\Pi}e^{-\tau\matr{L}}-\vectg{\pi}^\mathsf{T}\vectg{\pi}$.

\sloppy
In the case of asymmetric matrices, spectral clustering approaches usually rely on the singular vectors rather than on the eigenvectors to capture the structural asymmetries of a system\cite{Rohe2016}.
Similarily, the forward and backward clustering of our framework can be related to the clustering of the singular vectors of the transition matrix.
In the temporal case, the existence of a stationary distribution is not guaranteed, 
however, we have $\pvec(t_1) \Tmat(t_1,t_2) \Tmat^\text{inv}(t_2,t_1) = \pvec(t_1)$
and $\pvec(t_2) \Tmat_\text{inv}(t_2,t_1)\Tmat(t_1,t_2) = \pvec(t_2)$, 
i.e. $\pvec(t_1)$ and $\pvec(t_2)$ are left-eigenvectors of $\Tmat(t_1,t_2) \Tmat^\text{inv}(t_2,t_1)$ and 
$\Tmat^\text{inv}(t_2,t_1)\Tmat(t_1,t_2)$, respectively, with eigenvalue 1.
If the processes defined by  $\Tmat(t_1,t_2) \Tmat^\text{inv}(t_2,t_1)$ and 
$\Tmat^\text{inv}(t_2,t_1)\Tmat(t_1,t_2)$ are irreducibles, $\pvec(t_1)$ and $\pvec(t_2)$ are their respective stationary distributions.
Using the covariances of the forward and inverse backward flows (eqs. \ref{eq:autocov_forw} \& \ref{eq:autocov_back_inv}) is therefore a natural generalization of the Markov stability in the stationary case to the non-stationary case.
Moreover, the inverse transition matrix, $\Tmat^\text{inv}(t_2,t_1)=\Pmat(t_2)^{-1}\Tmat(t_1,t_2)^\mathsf{T}\Pmat(t_1)$, can be seen as the adjoint operator of $\Tmat(t_1,t_2)$ with respect to the inner product $\langle x,y \rangle_{t}=\sum_i x_i y_i / p_i(t)$, for which we have $\langle \vect{p}(t_1)\Tmat(t_1,t_2),\vect{p}(t_2)\rangle_{t_2}=\langle \vect{p}(t_1),\vect{p}(t_2)\Tmat^\text{inv}(t_2,t_1)\rangle_{t_1}$.
The vectors $\pvec(t_1)$ and $\pvec(t_2)$ are therefore singular vectors of the transition matrix with respect to this inner product.

\subsection*{Special cases of the random walk covariances in static networks}
\label{sec:special_cases}
Table S\ref{tab:relations} shows how modularity\cite{Newman2004}, directed-modularity\cite{Arenas2007,Kim2010}
and Markov stability\cite{Delvenne2010,Lambiotte2014} can be constructed from special
cases of the non-stationary clustered covariance from eq. (\ref{eq:clustered_autocov}).
Similarly, Tab. S\ref{tab:relations} shows that the clustering of static directed networks using the 
bibliographic coupling and co-citation matrices\cite{Satuluri2011} 
are special cases of the clustering with the forward and backward non-stationary covariances from 
eqs. (\ref{eq:autocov_forw}) and (\ref{eq:autocov_backw}), respectively.

\subsection*{Covariances of inverse processes}
\label{sec:inv_covar}

An alternative backward process than the one defined in eq. (\ref{eq:autocov_backw}) can be constructed by considering the inverse of the process that started at $t_1$ instead of the reversed evolution of the network.
In this case, the corresponding covariance is given by
\begin{align}
\matr{S}_{\text{back}}^{\textrm{inv}}(t_1,t) &= \matr{P}(t)\matr{T}^{\textrm{inv}}(t,t_1)\matr{T}(t_1,t) - \vect{p}(t)^\mathsf{T}\vect{p}(t) \notag\\
&= \matr{T}(t_1,t)^\mathsf{T}\matr{P}(t_1)\matr{T}(t_1,t) - \vect{p}(t)^\mathsf{T}\vect{p}(t).
  \label{eq:autocov_back_inv}
\end{align}

Similarily, the covariance of the inverse backward process of eq. \ref{eq:autocov_backw} is given by
\begin{align}
\matr{S}_{\text{forw}}^{\textrm{inv}}(t_2,t) &= \matr{P}(t)\matr{T}_\textrm{rev}^{\textrm{inv}}(t,t_2)\matr{T}_\textrm{rev}(t_2,t) - \vect{p}(t)^\mathsf{T}\vect{p}(t) \notag\\
&= \matr{T}(t_2,t)_\textrm{rev}^\mathsf{T}\matr{P}(t_1)\matr{T}_\textrm{rev}(t_2,t) - \vect{p}(t)^\mathsf{T}\vect{p}(t).
  \label{eq:autocov_forw_inv}
\end{align}

A difference between these two definitions is the choice of the initial condition, which is at $t_2$ for eq. \ref{eq:autocov_backw} and at $t_1$ for eq. \ref{eq:autocov_back_inv}.
The matrices $\Smat_\text{forw}(t_1, t)$ and $\Smat^\text{inv}_\text{back}(t_1, t)$ are both covariances of the same diffusion process that start at $t_1$ and evolve until $t>t_1$ while the matrices $\Smat_\text{forw}(t_1, t)$ and $\Smat_\text{back}(t_2, t)$ are the covariances of two different processes, the first starting at $t_1$ and evolving in the direction of time and the second starting at $t_2$ and evolving backward in time.
Here, we prefer to use $\Smat_\text{forw}(t_1, t)$ and $\Smat_\text{back}(t_2, t)$ for the general clustering of temporal networks between $t_1$ and $t_2$ using $\pvec(t_1)$ and $\pvec(t_2)$ as two uniform distributions. Using the inverse covariances (\ref{eq:autocov_back_inv}) and (\ref{eq:autocov_forw_inv}) may, for example, be preferred when studying a specific diffusion process.

\subsection*{Importance of early and late times on the optimal partitions}
\label{sec:time-weighting}

We consider a simple example of temporal network with \rev{eight} nodes ($N=8$) that initially forms two communities \rev{of four nodes each} and after a time $t^\star$ split to form four communities \rev{of two nodes}. The question we want to answer is how does the partition maximizing the forward flow stability (eq. \ref{eq:forw_int_stability}) changes as a function of $t^\star$ when the integration goes from $t_1=0 < t^\star$ until $t > t^\star$. 
The Laplacian matrix from $t=0$ until $t=t^\star$ is a matrix with four 4$\times$4 blocks. The off diagonal blocks are zero matrices and the diagonal blocks are two similar matrices given by
\begin{equation}
\matr{L}_A = 
\begin{pmatrix}
1 & -1/3 & -1/3 & -1/3 \\
-1/3 & 1 & -1/3 & -1/3  \\
-1/3 & -1/3 & 1 & -1/3 \\
-1/3 & -1/3 & -1/3 & 1 \\ 
\end{pmatrix}
.
\end{equation}
For times $t>t^\star$, the Laplacian has the same block structure with diagonal blocks given by
\begin{equation}
\matr{L}_B = 
\begin{pmatrix}
1 & -1  & 0 & 0 \\
-1 & 1 & 0 & 0  \\
0 & 0 & 1 & -1 \\
0 & 0 & -1 & 1 \\ 
\end{pmatrix}
.
\end{equation}
The two Laplacians are symmetric and commute. They are therefore simultaneously diagonalisable, i.e 
$\matr{L}_A = \matr{U \Lambda}_A\matr{ U}^\mathsf{T}$ and $\matr{L}_B = \matr{U \Lambda}_B\matr{ U}^\mathsf{T}$ with $\matr{\Lambda}_A = \text{diag}((4/3,4/3,4/3,0))$, $\matr{\Lambda}_B = \text{diag}((2,2,0,0))$ and where
\begin{equation}
\matr{U} = 
\begin{pmatrix}
0 & -1/{\sqrt{2}} & -1/2 & 1/2 \\
0 & 1/{\sqrt{2}} & -1/2 & 1/2  \\
-1/{\sqrt{2}} & 0 & 1/2 & 1/2 \\
1/{\sqrt{2}} & 0 & 1/2 & 1/2 \\ 
\end{pmatrix}
\end{equation}
is a unitary matrix.
The transition matrix is given by
\begin{equation}
    \Tmat(0,t) = \begin{cases}
    \matr{U}e^{-\lambda t \matr{\Lambda}_A}\matr{U}^\mathsf{T} = \matr{U}\matr{\Sigma}_A(t)\matr{U}^\mathsf{T} & \text{ if } 0\leq t \leq t^\star \\
    \matr{U}e^{-\lambda t^\star \matr{\Lambda}_A}e^{-\lambda (t-t^\star) \matr{\Lambda}_B}\matr{U}^\mathsf{T} =
    \matr{U}\matr{\Sigma}_A(t^\star)\matr{\Sigma}_B(t-t^\star)\matr{U}^\mathsf{T} & \text{ if } t > t^\star \\
    \end{cases}
    ,\label{eq:analy_trans_mat}
\end{equation}

where $\lambda$ is the random walk rate.
In the rest of this section, we use the notation $\matr{T}(t)$ as meaning $\matr{T}_1(0,t)$.
We can now calculate the forward covariance (eq. \ref{eq:autocov_forw}) taking $\vect{p}=\frac{1}{N}(1 1 1 1)$ as initial condition.
Note that $\vect{p}$ is a stationary state of the system, i.e. $\vect{p}=\vect{p}\matr{T}(t)$ $\forall t $ such that $t\geq 0$.
The forward covariance is given by $\matr{S}_{\text{forw}}(t) = \matr{P T}(t) \matr{P}^{-1} \Tmat(t)^\mathsf{T} \matr{P} - \vect{p}^\mathsf{T}\vect{p} = \frac{1}{N}\matr{T}(t)\matr{T}(t)^\mathsf{T} - \frac{1}{N^2}\overleftrightarrow{\matr{1}}$.
Or, using eq. (\ref{eq:analy_trans_mat})
\begin{equation}
\matr{S}_{\text{forw}}(t) = \begin{cases}
\frac{1}{N}\matr{U}\matr{\Sigma}_A^2(t)\matr{U}^\mathsf{T} - \frac{1}{N^2}\overleftrightarrow{\matr{1}}  & \text{ if } 0\leq t \leq t^\star \\
\frac{1}{N}\matr{U}\matr{\Sigma}_A^2(t^\star)\matr{\Sigma}_B^2(t-t^\star)\matr{U}^\mathsf{T} - \frac{1}{N^2}\overleftrightarrow{\matr{1}}  & \text{ if } t \geq t^\star \\
\end{cases}
.
\end{equation}

We find the forward flow stability by integrating $\matr{S}_{\text{forw}}(t')$ from 0 to $t$ and dividing by $t$. This yields the matrix 
\begin{equation}
\matr{F}_{\textrm{forw}}(t)=\begin{cases}
\frac{1}{tN} \matr{U} \int_0^t\matr{\Sigma}_A^2(t')dt'\matr{U}^\mathsf{T} - \frac{1}{N^2}\overleftrightarrow{\matr{1}} & \text{ if } 0\leq t \leq t^\star \\
\frac{t^\star}{t}\matr{F}_{\textrm{forw}}(t^\star) + \frac{1}{tN} \matr{U} \matr{\Sigma}_A^2(t^\star)\int_{t^\star}^t\matr{\Sigma}_B^2(t'-t^\star)dt'\matr{U}^\mathsf{T} - \frac{t-t^\star}{tN^2}\overleftrightarrow{\matr{1}} & \text{ if }t \geq t^\star \\
\end{cases}
,
\end{equation}
 where $\int_0^t\matr{\Sigma}_A^2(t')dt'= \text{diag}\big((
 \frac{3}{8\lambda}(1-e^{-\frac{8}{3}\lambda t}),\allowbreak  \frac{3}{8\lambda}(1-e^{-\frac{8}{3}\lambda t}),\allowbreak  \frac{3}{8\lambda}(1-e^{-\frac{8}{3}\lambda t}),\allowbreak t)\big)$ and 
 $\int_{t^\star}^t\matr{\Sigma}_B^2(t'-t^\star)
 dt'= \text{diag}\big(( \frac{1}{4\lambda}(1-e^{-4\lambda (t-t^\star)}),\allowbreak 
 \frac{1}{4\lambda}(1-e^{-4\lambda (t-t^\star)},\allowbreak 
 t-t^\star,\allowbreak t-t^\star)\big)$.
 The resulting matrix $\matr{F}_{\textrm{forw}}(t)$ is formed by four 4$\times$4 blocks. The off diagonal blocks are equal to $-\frac{1}{N^2}\overleftrightarrow{\matr{1}}$ and the diagonal blocks have the following form
 \begin{equation}
  \begin{pmatrix}
   F_{11}^\textrm{forw}(t) & F_{12}^\textrm{forw}(t) & F_{13}^\textrm{forw}(t) & F_{13}^\textrm{forw}(t)\\
   F_{12}^\textrm{forw}(t) & F_{11}^\textrm{forw}(t) & F_{13}^\textrm{forw}(t) & F_{13}^\textrm{forw}(t)\\
   F_{13}^\textrm{forw}(t) & F_{13}^\textrm{forw}(t) & F_{11}^\textrm{forw}(t) & F_{12}^\textrm{forw}(t)\\
   F_{13}^\textrm{forw}(t) & F_{13}^\textrm{forw}(t) & F_{12}^\textrm{forw}(t) & F_{11}^\textrm{forw}(t)\\
  \end{pmatrix}.
  \label{eq:flow_stab_matrix}
 \end{equation}
Moreover, for $ 0 \leq t\leq t^\star$, $F_{12}^\textrm{forw}(t) = F_{13}^\textrm{forw}(t)$.

The partition maximizing the forward flow stability groups together positive elements of $\matr{F}_{\textrm{forw}}(t)$ and avoids its negative elements. As the off diagonal blocks are always negative, regardless of $t$, the optimal partition depends on the signs of $F_{11}^\textrm{forw}(t), F_{12}^\textrm{forw}(t)$ and  $F_{13}^\textrm{forw}(t)$ as a function of t. 

We have
\begin{equation}
F_{11}^\textrm{forw}(t) = \begin{cases}
\frac{9}{32N\lambda t}\left(1 - e^{-\frac{8}{3}\lambda t}\right) + \frac{N-4}{4N^2}  & \text{ if } 0\leq t \leq t^\star \\
\frac{1}{32N\lambda t}\left(e^{-\frac{8}{3}\lambda t^\star}\left(8\lambda(t-t^\star)-4e^{-4\lambda(t-t^\star)}-5\right) + 9 + 8\lambda t\right)- \frac{1}{N^2} & \text{ if } t \geq t^\star \\
\end{cases},
\end{equation}

which is always positive, with $F_{11}^\textrm{forw}(0) = \frac{N-1}{N^2}$  and $F_{11}^\textrm{forw}(t) \rightarrow \frac{1}{4N^2}\left(N\left(e^{-\frac{8}{3}\lambda t^\star}+1\right)-4\right)$ as $ t\rightarrow\infty$.

We have
\begin{equation}
F_{12}^\textrm{forw}(t) = \begin{cases}
\frac{3}{32N\lambda t}\left(e^{-\frac{8}{3}\lambda t}-1\right) + \frac{N-4}{4N^2}  & \text{ if } 0\leq t \leq t^\star \\
\frac{1}{32N\lambda t}\left(e^{-\frac{8}{3}\lambda t^\star}\left(8\lambda(t-t^\star)+4e^{-4\lambda(t-t^\star)}-1\right) -3 + 8\lambda t\right)- \frac{1}{N^2} & \text{ if } t \geq t^\star \\
\end{cases},
\end{equation}
which is negative at $t=0$ with $F_{12}^\textrm{forw}(0)=-\frac{1}{N^2}$ and then increases monotonically.
At $t^\star$, $F_{12}^\textrm{forw}(t^\star)$ is positive only if $t^\star>\hat{t}_\textrm{f}$, where $\hat{t}_\textrm{f}$ is the time at which $F_{12}^\textrm{forw}(t)$ crosses the $x$-axis if this happens before $t^\star$. Its value is given by
\begin{equation}
 \hat{t}_\textrm{f} = \frac{3}{8\lambda}\left(\frac{N}{N-4}+W_0\left(-\frac{N}{N-4}e^{-\frac{N}{N-4}}\right)\right), 
 \label{eq:t_hat_f}
\end{equation}
where $W_0$ is the principal branch of the Lambert $W$ function. We see that $\hat{t}_\textrm{f}$ is made of two terms, a linear coefficient $\frac{1}{\lambda}$ and a constant term depending only on $N$ (which is fixed at $N=8$ here). By varying $\lambda$ one can therefore adjust $\hat{t}_\textrm{f}$ in order to make $F_{12}^\textrm{forw}(t)$ positive or negative for any $t$ such that $0<t\leq t^\star$.
As $t\rightarrow\infty$, $F_{12}^\textrm{forw}(t)\rightarrow\frac{1}{4N^2}\left(N\left(e^{-\frac{8}{3}\lambda t^\star}+1\right)-4\right)>0$ indicating that even if $F_{12}^\textrm{forw}(t^\star)$ is negative, $F_{12}^\textrm{forw}(t)$ eventually becomes positive.

For $F_{13}^\textrm{forw}(t)$, we have
\begin{equation}
F_{13}^\textrm{forw}(t) = \begin{cases}
F_{12}^\textrm{forw}(t)  & \text{ if } 0\leq t \leq t^\star \\
\frac{1}{32N\lambda t}\left(e^{-\frac{8}{3}\lambda t^\star}\left(3-8\lambda(t-t^\star)\right) -3 + 8\lambda t\right)- \frac{1}{N^2} & \text{ if } t \geq t^\star \\
\end{cases}.
\end{equation}
Similarly to $F_{12}^\textrm{forw}(t)$, $F_{13}^\textrm{forw}(t^\star)\allowbreak > 0$ only if $t^\star>\hat{t}_\textrm{f}$.
As $t \rightarrow \infty$, $F_{13}^\textrm{forw}(t)\allowbreak \rightarrow \frac{1}{4N^2}\left(N\left(1-e^{-\frac{8}{3}\lambda t^\star}\right)-4\right)$ 
which is positive only if $t^\star>t_{13}=\frac{3}{8\lambda}\ln{\frac{N}{N-4}}$ whose value can again be controlled by varying $\lambda$.

we remark that $F_{11}^\textrm{forw}(t) \geq F_{12}^\textrm{forw}(t) \geq F_{13}^\textrm{forw}(t)$, we have three possible configurations:
A) $F_{11}^\textrm{forw}(t) > 0$, $F_{12}^\textrm{forw}(t)> 0$ and $F_{13}^\textrm{forw}(t) > 0$: the optimal partition is composed of two communities of size 4;
B) $F_{11}^\textrm{forw}(t)> 0$, $F_{12}^\textrm{forw}(t)> 0$ and $F_{13}^\textrm{forw}(t) < 0$: the optimal partition is composed of four communities of size 2;
C) $F_{11}^\textrm{forw}(t)> 0$, $F_{12}^\textrm{forw}(t)<0$ and $F_{13}^\textrm{forw}(t)<0$: the optimal partition is composed of 8 singleton communities.

\begin{figure}[htp]
\centering
 \includegraphics[width=0.75\linewidth]{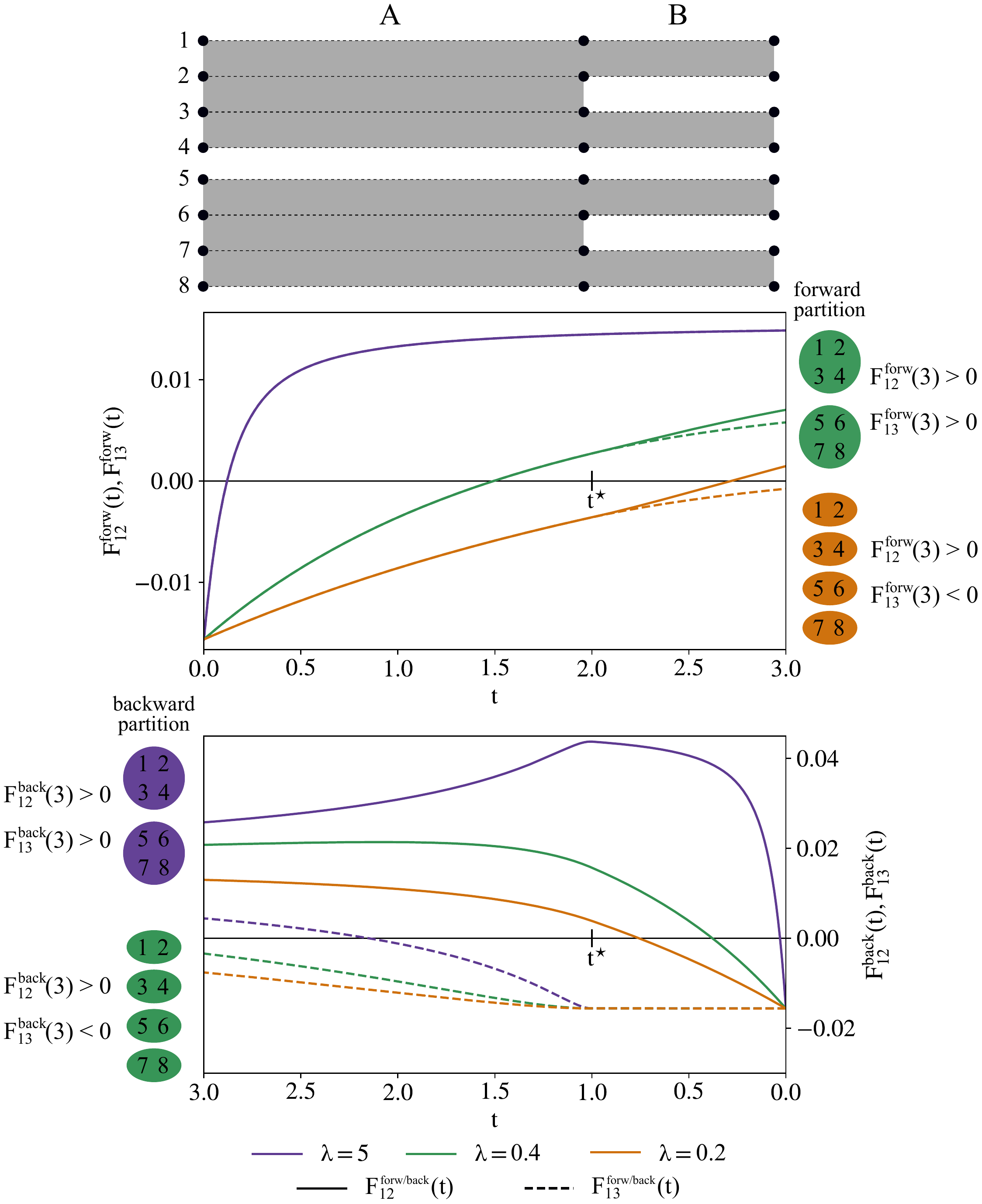}
 \caption{\textbf{Graph of the functions $F_{12}^\textrm{forw}(t)$ \rev{\& $F_{13}^\textrm{forw}(t)$} (top) and $F_{12}^\textrm{back}(t)$ \rev{\& $F_{13}^\textrm{back}(t)$} (bottom) for different values of the random walk rate $\lambda$.} The sign of these functions control whether the forward, respectively backward, optimal partitions take the form of the early times or later times.
 Here, the network splits from a structure, A, in \rev{2} communities to a structure, B, \rev{in four} communities at $t^\star=2$.
 By varying the value of $\lambda$, we can give give more importance to the structure at early times (fast diffusion) or to the structure at later times (slow diffusion).
 By considering a time interval that lasts 3 time units, the backward process starts at $t_2=3$ with a reverse time evolution (bottom). \rev{$F_{13}^\textrm{back}(t)$} can be positive only for values of the reverse time larger than $2$.
 We show three scenarios depending on the value of $\lambda$: 1) \rev{$\lambda=5$ (purple)}, the optimal forward and backward partitions have the form of A; 2) \rev{$\lambda=0.4$ (green)}, the optimal forward partition has the form of A and the optimal backward partition has the form of B; 3) \rev{$\lambda=0.2$ (yellow)}, the optimal forward and backward partitions have the form of B.}
 \label{fig:time_weighting}
\end{figure}

Noticing that $t_{13}<\hat{t}_\textrm{f}$, we therefore have three scenarios:
\begin{enumerate}
    \item[1)] $t^\star<t_{13}<\hat{t}_\textrm{f}$: $F_{12}^\textrm{forw}(t)$ and $F_{13}^\textrm{forw}(t)$ are negative at $t=t^\star$. 
    The switch to four communities happens too fast for the walkers to have time to explore the two community structure.
    After $t^\star$, $F_{13}^\textrm{forw}(t)$ remains negative while $F_{12}^\textrm{forw}(t)$ eventually becomes positive. For short intervals, the configuration C is optimal. For longer intervals, the configuration B becomes optimal. This change can be controlled by varying $\lambda$.
    \item[2)] $t_{13}<t^\star<\hat{t}_\textrm{f}$: As before, $F_{12}^\textrm{forw}(t)$ and $F_{13}^\textrm{forw}(t)$ are negative at $t=t^\star$.
    While the walkers have not fully explored the two communities structure at $t^\star$, they have sufficiently done so that after some time  $F_{12}^\textrm{forw}(t)$ and $F_{13}^\textrm{forw}(t)$ both become positive and the optimal forward partition is given by configuration A.
     \item[3)]$t_{13}<\hat{t}_\textrm{f}<t^\star$: $F_{12}^\textrm{forw}(t^\star>$ and $F_{13}^\textrm{forw}(t^\star)$ are already positive at $t=t^\star$ and remain positive afterward. The walkers have already fully explored the two community structure before $t^\star$ and the optimal forward partition remains given by configuration A.

\end{enumerate}
 Figure S\ref{fig:time_weighting} shows a graph of $F_{12}^\textrm{forw}(t)$ and $F_{13}^\textrm{forw}(t)$ for different values of $\lambda$. We see that the importance of early or late times on the forward partition can be controlled by varying the value of the random walk rate.
Indeed, the three conditions corresponding to the three scenarios are expressed as inequalities between $\lambda t^\star$ and constants that depends only on the structure of the network.

Considering the backward evolution by reversing time, the system starts in the configuration with four communities at $t=0$ until $t^\star$ and then forms the structure in two communities.
For the backward case, we have
\begin{equation}
    \Tmat(t)_\textrm{rev} = \begin{cases}
     \matr{U}\matr{\Sigma}_B(t)\matr{U}^\mathsf{T} & \text{ if } 0\leq t \leq t^\star \\
    \matr{U}\matr{\Sigma}_B(t^\star)\matr{\Sigma}_A(t-t^\star)\matr{U}^\mathsf{T} & \text{ if } t > t^\star \\
    \end{cases}
    ,\label{eq:analy_trans_mat_back}
\end{equation}
\begin{equation}
\matr{S}_{\text{back}}(t) = \begin{cases}
\frac{1}{N}\matr{U}\matr{\Sigma}_B^2(t)\matr{U}^\mathsf{T} - \frac{1}{N^2}\overleftrightarrow{\matr{1}}  & \text{ if } 0\leq t \leq t^\star \\
\frac{1}{N}\matr{U}\matr{\Sigma}_B^2(t^\star)\matr{\Sigma}_A^2(t-t^\star)\matr{U}^\mathsf{T} - \frac{1}{N^2}\overleftrightarrow{\matr{1}}  & \text{ if } t \geq t^\star \\
\end{cases}
,
\end{equation}
\begin{equation}
\matr{F}_{\textrm{back}}(t)=\begin{cases}
\frac{1}{tN} \matr{U} \int_0^t\matr{\Sigma}_B^2(t')dt'\matr{U}^\mathsf{T} - \frac{1}{N^2}\overleftrightarrow{\matr{1}} & \text{ if } 0\leq t \leq t^\star \\
\frac{t^\star}{t}\matr{F}_{\textrm{back}}(t^\star) + \frac{1}{tN} \matr{U} \matr{\Sigma}_B^2(t^\star)\int_{t^\star}^t\matr{\Sigma}_A^2(t'-t^\star)dt'\matr{U}^\mathsf{T} - \frac{t-t^\star}{tN^2}\overleftrightarrow{\matr{1}} & \text{ if }t \geq t^\star \\
\end{cases}.
\end{equation}

The backward flow stability matrix has the same structure than in the forward case (eq. \ref{eq:flow_stab_matrix}), but with 
\begin{equation}
F_{11}^\textrm{back}(t) = \begin{cases}
\frac{1}{8N\lambda t}\left(1 - e^{-4\lambda t}\right) + \frac{N-2}{2N^2}  & \text{ if } 0\leq t \leq t^\star \\
\parbox{8cm}{$\frac{1}{32N\lambda t}(e^{-4\lambda t^\star}(2-6e^{-\frac{8}{3}\lambda(t-t^\star)}-3e^{-\frac{4}{3}\lambda(2t-5t^\star)}) + 7 + 8\lambda (t+t^\star))- \frac{1}{N^2}$} & \text{ if } t \geq t^\star \\
\end{cases},
\end{equation}
which is always positive with a value of $\frac{N-1}{N^2}$ at $t=0$ and $\frac{N-4}{4N^2}$ as $t\rightarrow\infty$.
\begin{equation}
F_{12}^\textrm{back}(t) = \begin{cases}
\frac{1}{8N\lambda t}\left(e^{-4\lambda t}-1\right) + \frac{N-2}{2N^2}  & \text{ if } 0\leq t \leq t^\star \\
\parbox{8cm}{$\frac{1}{32N\lambda t}(e^{-4\lambda t^\star}(6e^{-\frac{8}{3}\lambda(t-t^\star)}-2-3e^{-\frac{4}{3}\lambda(2t-5t^\star)}) -1 + 8\lambda (t+t^\star))- \frac{1}{N^2}$} & \text{ if } t \geq t^\star \\
\end{cases},
\end{equation}
which starts with a negative value of $-\frac{1}{N^2}$ at $t=0$ and increases until $t=t^\star$. $F_{12}^\textrm{back}(t^\star)$ is positive only if $t^\star>\hat{t}_\textrm{b}$ where
\begin{equation}
    \hat{t}_\textrm{b}= \frac{1}{4\lambda}\left(\frac{N}{N-2}+W_0\left(-\frac{N}{N-2}e^{-\frac{N}{N-2}}\right)\right).
    \label{eq:t_hat_b}
\end{equation}
As $t\rightarrow\infty$, $F_{12}^\textrm{back}(t)\rightarrow\frac{N-4}{4N^2}$, i.e. $F_{12}^\textrm{back}(t)$ eventually becomes positive if it was not the case at $t^\star$ or stays positive otherwise.
\begin{equation}
F_{13}^\textrm{back}(t) = \begin{cases}
-\frac{1}{N^2}  & \text{ if } 0\leq t \leq t^\star \\
\frac{1}{32N\lambda t}\left(3e^{-\frac{8}{3}\lambda(t-t^\star)}+8\lambda(t-t^\star)\right)- \frac{1}{N^2} & \text{ if } t \geq t^\star \\
\end{cases},
\end{equation}
which is negative until $t=t^\star$ and then increases monotonically, eventually becomes positive and reaches a value of $\frac{N-4}{4N^2}$ as $t\rightarrow\infty$.
The time at which $F_{13}^\textrm{back}(t)$ becomes positive is given by
\begin{equation}
    \hat{t}'_\textrm{b}= \frac{1}{8\lambda(N-4)}\left(3N+8\lambda Nt^\star+(3N-12)W_0\left(-\frac{N}{N-4}e^{-\frac{3N+32\lambda t^\star}{3N-12}}\right)\right).
\end{equation}
Contrary to $\hat{t}_\textrm{f}$ (eq. \ref{eq:t_hat_f}) and $\hat{t}_\textrm{b}$ (eq. \ref{eq:t_hat_b}) that both tend to zero as the speed of the walkers is increased, $\hat{t}'_\textrm{b}$ tends to $2t^\star$ as $\lambda\rightarrow\infty$.
This indicates that $F_{13}^{\textrm{back}}(t)$ can become positive only for times larger than $2t^\star$. The importance of early and late time can therefore also be controlled with $\lambda$, however there is a limit on the possibility of detecting the first structure (two communities) if it lasts for a shorter time than the second structure (four communities). 
This is due to the fact that the second structure (configuration B) is composed of smaller communities. When looking at the network evolution from the point of view of the backward partition, i.e. from the end of the interval backward in time, the vision of the first structure can be obstructed by the second smaller structures. 
In this case, the first structure can be captured by the forward partition.

As for the forward partition, for the backward case we have $F_{11}^\textrm{back}(t) \geq F_{12}^\textrm{back}(t) \geq F_{13}^\textrm{back}(t)$ for $t>0$. We therefore have the following scenarios: 
1) $t<\hat{t}'_\textrm{b}$, $F_{11}^\textrm{back}(t)$ is positive and $F_{13}^\textrm{back}(t)$ is negative. The sign of $F_{12}^\textrm{back}(t)$ is controlled by the RW rate $\lambda$. For slow RWs, configuration C is optimal (singletons communities). For fast RWs, configuration B is optimal (four communities).
2) $t>\hat{t}'_\textrm{b}$,  $F_{11}^\textrm{back}(t)$ is positive and the signs of $F_{12}^\textrm{back}(t)$ and $F_{13}^\textrm{back}(t)$ is controlled by $\lambda$. As the RW rate increases, the optimal backward partition changes from configurations C to B and finally A.

Figure S\ref{fig:time_weighting} shows a graph of $F_{12}^{\textrm{back}}(t)$ and $F_{13}^{\textrm{back}}(t)$ for different values of $\lambda$ showing that the evolution of the network can be captured by combining the solutions of the optimal forward and backward partitions for different values of $\lambda$.

\newpage

\begin{figure}[htp]
\centering
 \includegraphics[width=0.9\linewidth]{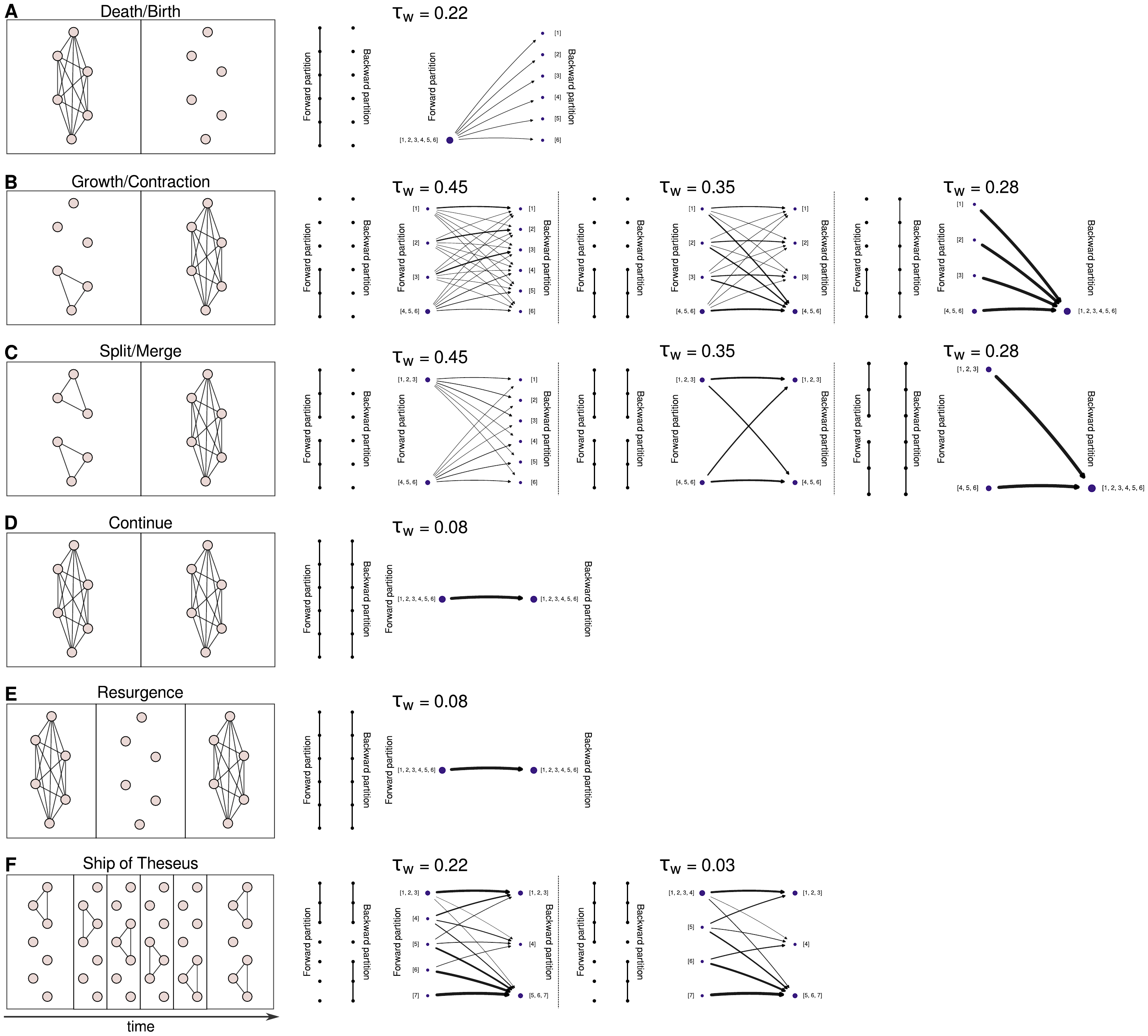}
 \caption{\textbf{Results of the flow stability community detection applied to toy examples of dynamic community events.} We reproduce the community events from Ref. \cite{Rossetti2018} with the addition of the Ship of Theseus\cite{Cazabet2021}.
 For each event, we show the schematic evolution on the left and the flow stability results on the right.
 Note that our framework does not distinguish between a node that is absent or a node that is inactive. Such nodes are usually clustered in singleton communities.
 We show the results for several values of the resolution (waiting time $\tau_w$) when several non trivial solutions exists.
 For each result, we represent the partitions in two manners: 1) the nodes as dots and the forward and backward communities as lines joining the dots, 2) as a bipartite graph where the nodes represents the communities and the edges represents the probability transitions from forward to backward communities of the random walk. The death/birth (\textbf{A}), growth/contraction (\textbf{B}), split/merge (\textbf{C}) and continue (\textbf{D}) events are well detected by our method. However, the flow stability is unable to distinguish the resurgence event (\textbf{E}) from the continue event as the absence of all connections in the middle results in an unchanged diffusion. To distinguish such situations, one has to split the time window in two resulting in a sequence of 'death' and 'birth' (\textbf{A}).
 In the ship of Theseus (\textbf{F}), the initial and final states are captures by the forward and backward partitions. To capture the dynamics between those states one needs to look at the transitions probabilities: there are non-zero probabilities to go from the initial ship to the two final ships, however there is a zero probability to go from nodes of the bottom ship at the beginning to the top ship at the end. We understand that the two final ships are linked to the initial ship.}
 \label{fig:community_events}
\end{figure}

\begin{figure}[htbp]
    \centering
    \includegraphics[width=0.8\linewidth]{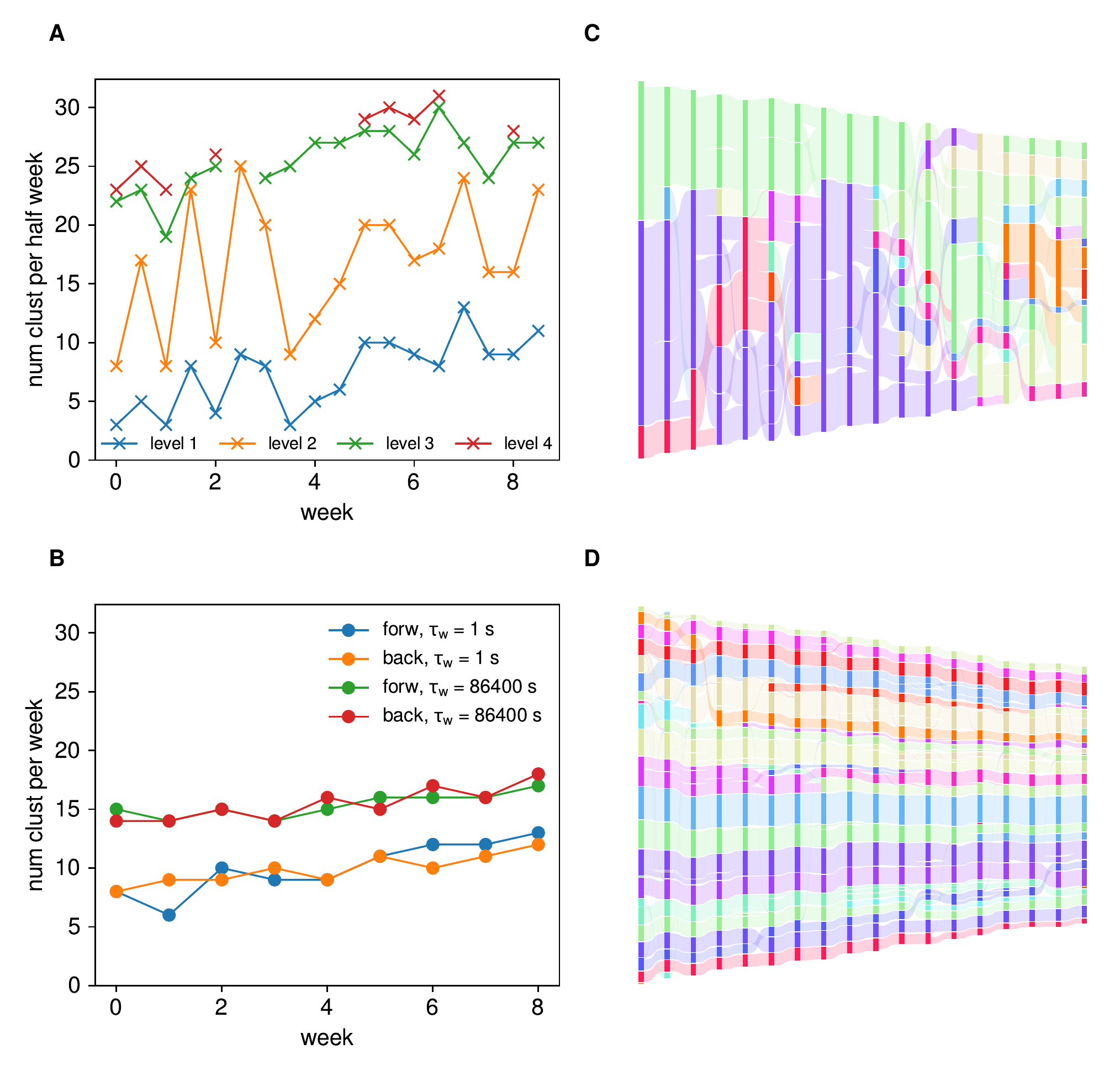}
    \caption{\textbf{Community detection of the free-ranging house mice contact network using a series of static networks with an aggregation window of a half week.}
    The hierarchical infomap algorithm\cite{Rosvall2011} is used on each time slice and the evolution of the communities is tracked using the method developed in Ref.\cite{Liechti2019} with a history parameter of 8 time points.
    The infomap algorithm is run with the default parameters for the hierarchical case.
    (\textbf{A}) Number of communities per time slice found by infomap for different hierarchical levels. An issue with this approach is that the method does not necessarily find the same number of hierarchical levels at each time slice which makes the comparison from slice to slice not clear.
    (\textbf{B}) Number of communities per week found with the flow stability for the two resolutions shown in the main manuscript. As the resolution parameter of the flow stability can be interpreted as a physical quantity (characteristic waiting time of the random walk) the comparison between different time slices is done in a principled manner. We see that, contrary to the case in (\textbf{A}), the number of clusters per week varies more smoothly with the flow stability.
    (\textbf{C} \& \textbf{D}) Result of the tracking of communities found with infomap for the coarsest level (i.e. level 1) and the finest level (i.e. 2, 3 or 4 depending on the time slice), respectively.}
    \label{fig:mice_alluv_compa}
\end{figure}

\begin{figure}[htbp]
    \centering
    \includegraphics[width=0.65\linewidth]{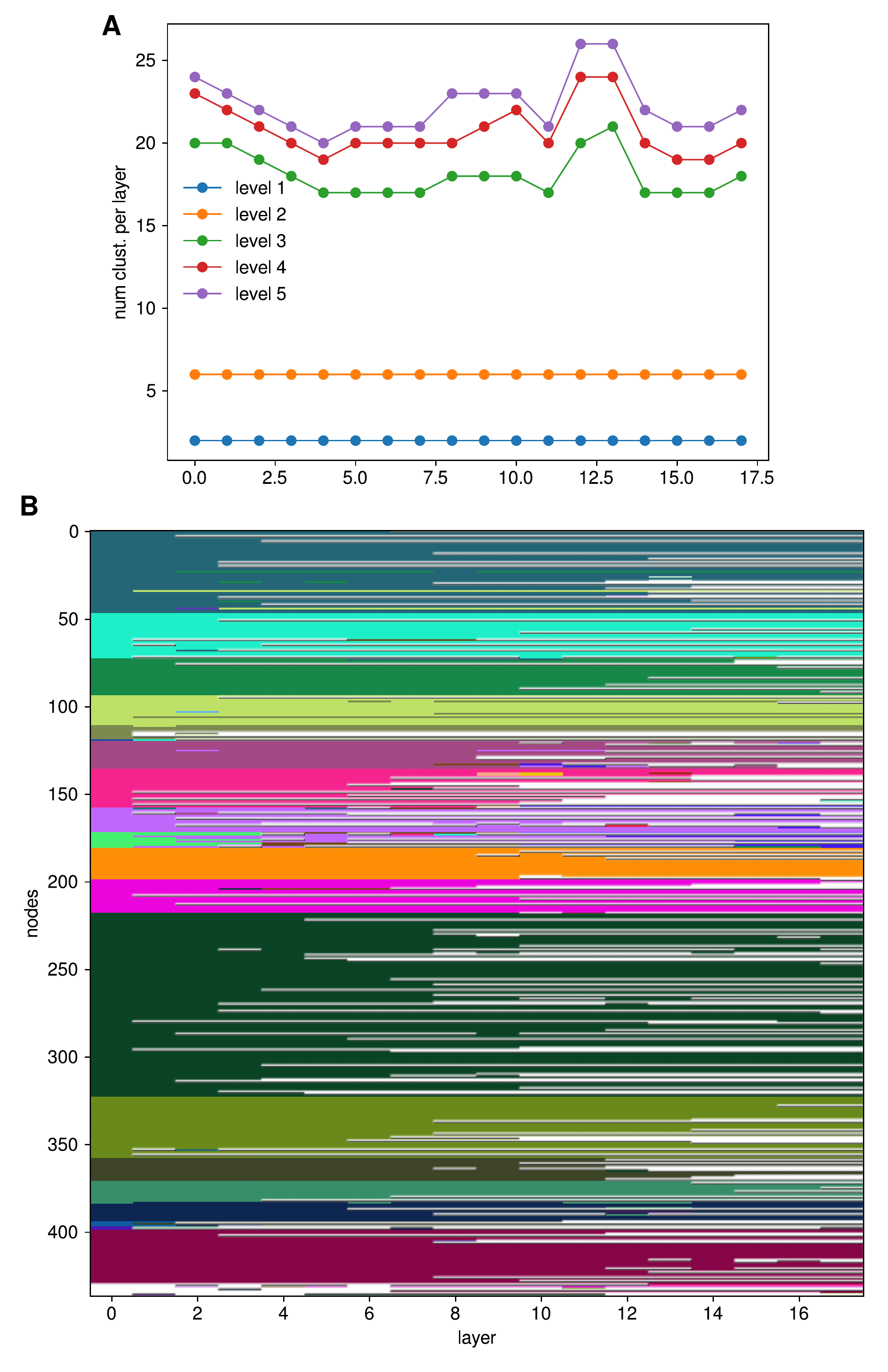}
    \caption{\textbf{Hierarchical multilayer community detection with the Infomap algorithm applied to the free-ranging house mice contact network.}
    A multilayer representation of this dataset is constructed by aggregating the activity in static networks over half-week time windows.
    The Infomap algorithm\cite{DeDomenico2015} is run with the parameters
    \texttt{flow\_model='undirected'},
    \texttt{multilayer\_relax\_by\_jsd} for neighborhood flow coupling for temporal networks\cite{Aslak2018},
    \texttt{multilayer\_relax\_limit=1} limiting the RW to jump only to neighboring layers in order to encode the temporal ordering,
    and with a value \texttt{multilayer\_relax\_rate} of 0.001.
    (\textbf{A}) Number of communities per layer (half-week) found by the multilayer Infomap algorithm at different hierarchical levels.
    (\textbf{B}) Multilayer partition at hierarchical level 3.
    Here, Infomap find 5 scales of communities considering all time slices simultaneously, however the communities found are all elongated in time and the dynamics of communities splitting found by the flow stability is not recovered.}
    \label{fig:mice_multilay}
\end{figure}

\begin{figure}[htbp]
\centering
 \includegraphics[width=\linewidth]{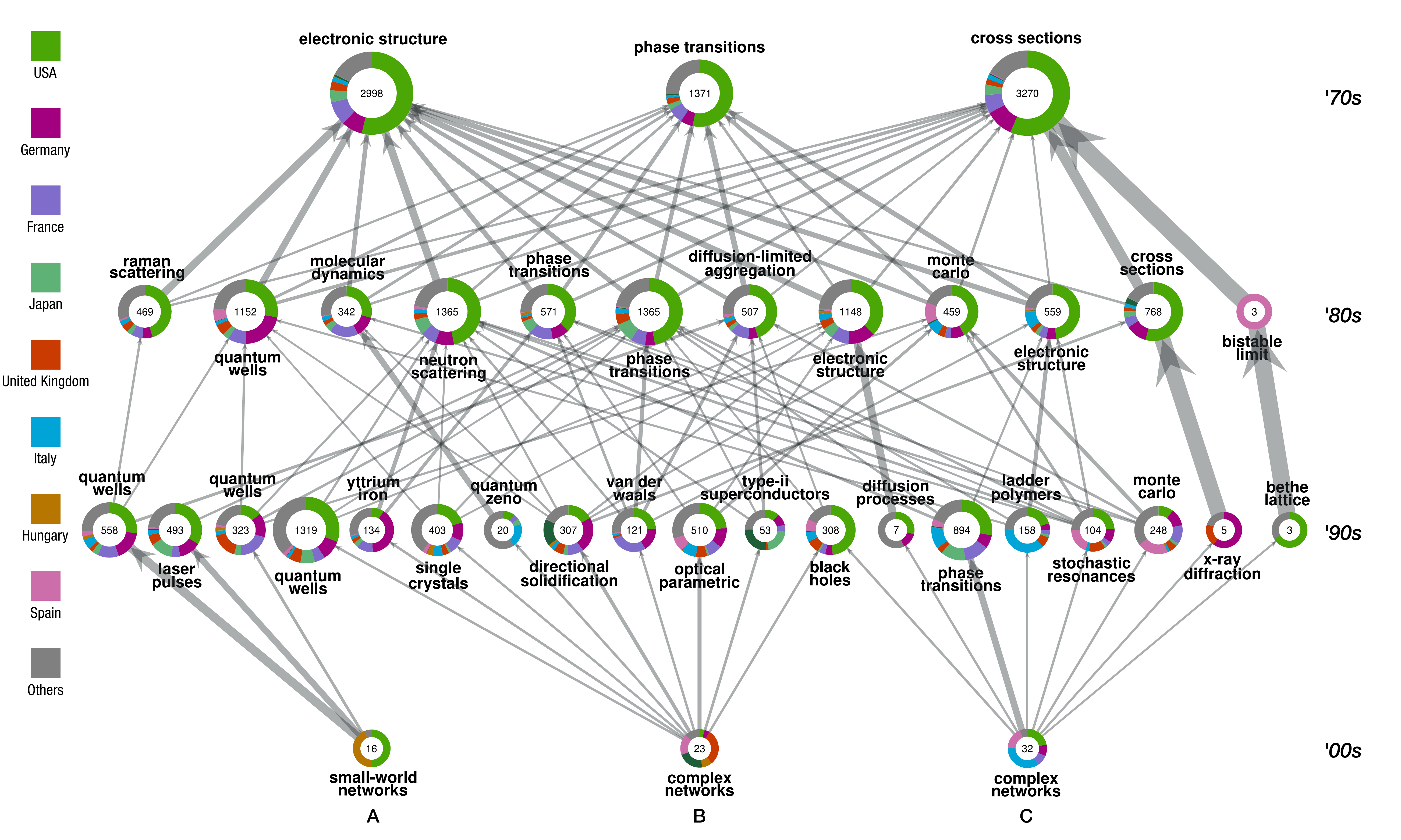}
 \caption{\textbf{Influential communities of authors of articles published in the APS journals for three communities of network scientists in the 2000s.}
 Each node represents a community and its size is indicated in the center.
 The colors represent the distribution of countries inside each community where each author is associated to the country most often associated with their affiliations.
 The pair of words next to each node indicate one of the most frequent pair of words of all the titles of the articles belonging to the community.
 Arrows between the communities represents probability transitions  ($>5\%$) from community to community of the diffusive process starting in 2010 and finishing in 1970.}
 \label{fig:complenet_flow_countries}
\end{figure}

\clearpage
\newpage

\begin{table}
\footnotesize
    \centering
    \begin{tabular}{ccccccc}
\toprule
{}             &  \multicolumn{2}{c}{static aggregation} & \multicolumn{2}{c}{NMI with fast flow stability} & \multicolumn{2}{c}{NMI with slow flow stability} \\
{slice length} &             from start &  from end &  forward &  backward &  forward &  backward \\
\midrule
0.1 &  \{a, b, c\}, \{d\} &  \{a\}, \{b, c, d\} &       1.00 &       1.00 &       0.54 &       0.31 \\
0.2 &  \{a, b, c\}, \{d\} &  \{a\}, \{b, c, d\} &       1.00 &       1.00 &       0.54 &       0.31 \\
0.3 &  \{a, b, c\}, \{d\} &  \{a, d\}, \{b, c\} &       1.00 &       0.31 &       0.54 &       1.00 \\
0.4 &  \{a, b, c\}, \{d\} &  \{a, d\}, \{b, c\} &       1.00 &       0.31 &       0.54 &       1.00 \\
0.5 &  \{a, d\}, \{b, c\} &  \{a, d\}, \{b, c\} &       0.31 &       0.31 &       0.67 &       1.00 \\
0.6 &  \{a, d\}, \{b, c\} &  \{a, d\}, \{b, c\} &       0.31 &       0.31 &       0.67 &       1.00 \\
0.7 &  \{a, d\}, \{b, c\} &  \{a, d\}, \{b, c\} &       0.31 &       0.31 &       0.67 &       1.00 \\
0.8 &  \{a, d\}, \{b, c\} &  \{a, d\}, \{b, c\} &       0.31 &       0.31 &       0.67 &       1.00 \\
0.9 &  \{a, d\}, \{b, c\} &  \{a, d\}, \{b, c\} &       0.31 &       0.31 &       0.67 &       1.00 \\
1.0 &  \{a, d\}, \{b, c\} &  \{a, d\}, \{b, c\} &       0.31 &       0.31 &       0.67 &       1.00 \\
\bottomrule
\end{tabular}
    \caption{\rev{Comparison between static partitions with different aggregation length and the flow stability partitions from the example in Fig. 1. The first column shows the slice lengths expressed as a ratio of the total length. The second and third columns show the partitions found by optimizing modularity on the networks found by aggregating from the beginning and end of the network with increasing slice lengths.
    The remaining columns show the value of the Normalized Mutual Information computed between the initial and final static partitions and the forward and backward flow stability partitions, respectively, for the case of the fast and slow diffusion of Fig. 1. Similar results are obtained when self-loops with weight corresponding to the inactivity time of nodes are added during the aggregation. We see that the NMI with the slow forward partition is never equal to one, indicating that the static aggregations cannot fully reproduce the results of the flow stability.}}
    \label{tab:comparison_slice_len}
\end{table}

\begin{table}
\footnotesize
\centering
\begin{tabular}{lcccccccccc}
\toprule
Forward   &            1 &            2 &            3 &            4 &            5 &            6 &            7 &            8 &            9 &           10 \\
\midrule
size       &           49 &           50 &           46 &           22 &           24 &           47 &            1 &            1 &            1 &            1 \\
\multirow{2}{*}{$\bar{T}_{\text{first}}$}   &        Day 1 &        Day 1 &        Day 1 &        Day 1 &        Day 1 &        Day 1 &        Day 2 &        Day 2 &        Day 2 &        Day 2 \\
&        09:00 &        09:32 &        08:56 &        09:04 &        09:56 &        09:04 &        08:43 &        08:41 &        08:42 &        08:42 \\
\bottomrule
\end{tabular}
\vspace{0.5cm}

\begin{tabular}{lcccccccccc}
\toprule
Backward         &            1 &            2 &            3 &            4 &            5 &            6 &            7 &            8 &            9 &           10 \\
\midrule
size       &           24 &           26 &           51 &           46 &           24 &           22 &           46 &            1 &            1 &            1 \\
\multirow{2}{*}{$\bar{T}_{\text{last}}$}     &        Day 2 &        Day 2 &        Day 2 &        Day 2 &        Day 2 &        Day 2 &        Day 2 &        Day 1 &        Day 1 &        Day 1 \\
   &        17:09 &        17:00 &        17:03 &        17:03 &        13:21 &        11:58 &        17:00 &        11:59 &        17:08 &        17:04 \\
\bottomrule
\end{tabular}

\caption{Size, average time of the first ($\bar{T}_{\text{first}}$) and last ($\bar{T}_{\text{last}}$) 
contact for each cluster of the forward and backward flow stability partitions at scale $\tau_w=1$\,h.}
\label{tab:part_1h}
\end{table}

\begin{table}
\footnotesize
\centering
 \begin{tabular}{lccccccccccc}
\toprule
Forward        &            1 &            2 &            3 &            4 &            5 &            6 &            7 &            8 &            9 &           10 &           11 \\
\midrule
size       &          114 &           52 &           67 &            1 &            1 &            1 &            1 &            1 &            2 &            1 &            1 \\
\multirow{2}{*}{$\bar{T}_{\text{first}}$}   &        Day 1 &        Day 1 &        Day 1 &        Day 1 &        Day 2 &        Day 2 &        Day 2 &        Day 2 &        Day 1 &        Day 2 &        Day 2 \\
 &        08:55 &        09:04 &        08:55 &        13:21 &        09:40 &        08:42 &        08:42 &        08:40 &        14:17 &        08:41 &        08:43 \\
\bottomrule
\end{tabular}

\begin{tabular}{lcccccccccccc}
\toprule
Backward         &            1 &            2 &            3 &            4 &            5 &            6 &            7 &            8 &            9 &           10 &           11 &           12 \\
\midrule
size       &          141 &           26 &           10 &           21 &           23 &            1 &            1 &            1 &            1 &           15 &            1 &            1 \\
\multirow{2}{*}{$\bar{T}_{\text{last}}$}    &        Day 2 &        Day 2 &        Day 2 &        Day 2 &        Day 2 &        Day 1 &        Day 2 &        Day 2 &        Day 1 &        Day 2 &        Day 1 &        Day 1 \\
   &        17:07 &        17:08 &        11:58 &        12:52 &        17:02 &        17:04 &        12:18 &        13:44 &        17:08 &        14:18 &        11:59 &        17:10 \\
\bottomrule
\end{tabular}

\caption{Size, average time of the first ($\bar{T}_{\text{first}}$) and last ($\bar{T}_{\text{last}}$) 
contact for each cluster of the forward and backward flow stability partitions at scale $\tau_w=63$\,s.}
\label{tab:part_63s}
\end{table}

\begin{sidewaystable}
 \scriptsize
 \begin{tabular}{lccccc}
		      &Probability	 	& Transition  		& Probability 		& Covariance	& Partition \\
		      &density 		 	& matrix: 		& density  		& matrix: 		& quality  \\
		      &at $t_1$: $\vect{p}(t_1)$& $\matr{T}(t_1,t_2)$	& at $t_2$: $\vect{p}(t_2)$ & 	$\matr{S}$	& function: \\
\toprule
1) \textbf{General}	& \multirow{3}{*}{$\vect{p}_1$}  & \multirow{3}{*}{Eq. (\ref{eq:temporal_trans_mat})}& \multirow{3}{*}{$\vect{p}_2=\vect{p}_1\matr{T}(t_1,t_2)$} & \multirow{3}{*}{\minitab[c]{$\matr{P}_1\matr{T}(t_1,t_2)-\vect{p}_1^\mathsf{T}\vect{p}_2$}} &  \\ 
non-stationary	&   				 	& 				&					& 		 &  \\
		&   				 	& 				&					& 		 &  \\
\hline
2)\textbf{Markov Stability}:& \multirow{3}{*}{$\vect{\pi}_i=\frac{k_i}{2M}, \forall i$}  & \multirow{3}{*}{\minitab[c]{$\tau = t_2 - t_1$\\$e^{-\tau\matr{L}}$}}& \multirow{3}{*}{$\vect{\pi}_i=\frac{k_i}{2M}, \forall i$} & \multirow{3}{*}{$\matr{\Pi}e^{-\tau\matr{L}}-\vectg{\pi}^\mathsf{T}\vectg{\pi}$} & \multirow{3}{*}{$\textrm{trace}\left[\matr{H}^\mathsf{T}\matr{S}(\tau)\matr{H}\right]$} \\
static			&   & 	& &  &  \\ 
 undirected&   		& 	& &  &  \\ 
\hline
3) \textbf{Modularity}:		& \multirow{3}{*}{$\vect{\pi}_i=\frac{k_i}{2M}, \forall i$}  & discrete-time	& \multirow{3}{*}{$\vect{\pi}_i=\frac{k_i}{2M}, \forall i$} & \multirow{3}{*}{$\matr{\Pi}\matr{D}^{-1}\matr{A}-\vectg{\pi}^\mathsf{T}\vectg{\pi}$} & \multirow{3}{*}{$\frac{1}{2 M}\sum_{ij}\left(A_{i j}-\frac{k_{i} k_{j}}{2 M}\right) \delta\left(c_{i}, c_{j}\right)$} \\
static 				&   						  & one step											& &  &  \\ 
undirected      		&   						  & $\matr{D}^{-1}\matr{A}$											& &  &  \\ 
\hline
4) \textbf{Dir. Modularity}:	& \multirow{3}{*}{$p_{1,i}=\frac{k_i^{\textrm{o}}}{M}, \forall i$}  & discrete-time& \multirow{3}{*}{$p_{2,i}=\frac{k_i^{\textrm{i}}}{M}, \forall i$} & \multirow{3}{*}{$\frac{1}{M}\matr{D}_{\textrm{o}}\matr{D}_{\textrm{o}}^{-1}\matr{A}-\vect{p}_1^\mathsf{T}\vect{p}_2$} & \multirow{3}{*}{$\frac{1}{M}\sum_{ij}\left(A_{i j}-\frac{k^\textrm{o}_{i} k^\textrm{i}_{j}}{M}\right) \delta\left(c_{i}, c_{j}\right)$} \\
Static 			&							   & one step	& &  &  \\ 
directed		&							   & $\matr{D}_{\textrm{o}}^{-1}\matr{A}$	& &  &  \\ 
\toprule
5) \textbf{Forward}:	& \multirow{3}{*}{$\vect{p_1}$}  & \multirow{3}{*}{Eq. (\ref{eq:temporal_trans_mat})}& \multirow{3}{*}{$\vect{p}_2=\vect{p}_1\matr{T}(t_1,t_2)$} & \multirow{3}{*}{\minitab[c]{$\matr{P}_1\matr{T}(t_1,t_2)\matr{P}_2^{-1}\matr{T}(t_1,t_2)^\mathsf{T}\matr{P}_1$\\$ - \vect{p}_1^\mathsf{T}\vect{p}_1$}} & \multirow{3}{*}{} \\ 
general &   & 	& &  &  \\ 
non-stationary 	&   & 	& &  &  \\ 
\hline
6) \textbf{Backward}:	& \multirow{3}{*}{$\vect{p}_1=\vect{p}_2\matr{T}_\text{rev}(t_2,t_1)$}  & \multirow{3}{*}{\minitab[c]{Eq. (\ref{eq:temporal_trans_mat})\\time-reversed\\$\matr{T}_\text{rev}(t_2,t_1)$}}& \multirow{3}{*}{$\vect{p}_2$} & \multirow{3}{*}{\minitab[c]{$\matr{P}_2\matr{T}_\text{rev}(t_2,t_1)\matr{P}_1^{-1}\matr{T}_\text{rev}(t_2,t_1)^\mathsf{T}\matr{P}_2$\\$ - \vect{p}_2^\mathsf{T}\vect{p}_2$}} & \multirow{3}{*}{} \\ 
general &   & 	& &  &  \\ 
non-stationary 	&   & 	& &  &  \\ 
\hline
7) \textbf{Bib. coupling}:	& \multirow{3}{*}{$p_{1,i}=\frac{k_i^{\textrm{o}}}{M}, \forall i$}  & discrete-time& \multirow{3}{*}{$p_{2,i}=\frac{k_i^{\textrm{i}}}{M}, \forall i$} & \multirow{3}{*}{$\frac{1}{M}\matr{A}\matr{D}_{\textrm{i}}^{-1}\matr{A}^\mathsf{T}-\vect{p}_1^\mathsf{T}\vect{p}_1$} & \multirow{3}{*}{$\frac{1}{M}\sum_{ij}\left(\sum_k\frac{A_{i k}A_{k j}}{k_k^{\textrm{i}}}-\frac{k^\textrm{o}_{i} k^\textrm{o}_{j}}{M}\right) \delta\left(c_{i}, c_{j}\right)$} \\
Static 			&							   & one step	& &  &  \\ 
directed		&							   & $\matr{D}_{\textrm{o}}^{-1}\matr{A}$	& &  &  \\ 
\hline
8) \textbf{Co-citation}:& \multirow{3}{*}{$p_{1,i}=\frac{k_i^{\textrm{o}}}{M}, \forall i$}  & reversed& \multirow{3}{*}{$p_{2,i}=\frac{k_i^{\textrm{i}}}{M}, \forall i$} & \multirow{3}{*}{$\frac{1}{M}\matr{A}^\mathsf{T}\matr{D}_{\textrm{o}}^{-1}\matr{A}-\vect{p}_2^\mathsf{T}\vect{p}_2$} & \multirow{3}{*}{$\frac{1}{M}\sum_{ij}\left(\sum_k\frac{A_{k i}A_{j k}}{k_k^{\textrm{o}}}-\frac{k^\textrm{i}_{i} k^\textrm{i}_{j}}{M}\right) \delta\left(c_{i}, c_{j}\right)$} \\
Static 			&							   & one step	& &  &  \\ 
directed		&							   & $\matr{D}_{\textrm{i}}^{-1}\matr{A}^\mathsf{T}$	& &  &  \\ 
\end{tabular}
\caption{Relation between non-stationary random walk covariance matrices and known partition quality functions.}
\label{tab:relations}
\end{sidewaystable}

\begin{table}[h]
    \centering
    \begin{tabular}{l}
         complex network \\
         scale-free network \\ 
         small-world network\\
         weighted network\\
         directed network\\
         growing network\\
         evolving network\\
    \end{tabular}
    \caption{Keywords used to find authors of articles about complex networks in the American Physical Society journals. The titles and abstracts of articles published between 2000 and 2010 were searched.}
    \label{tab:complenet_keywords}
\end{table}

\begin{sidewaystable}
\footnotesize
\begin{tabular}{llp{2cm}llp{2cm}lll}

\toprule
                     Community A &         &          &
                     Community B &         &          &  
                     Community C &         &         \\
\midrule
        A. L. Barabási &      USA &  PhysRevE &          Chiu Fan Lee &  United Kingdom &  PhysRevE &      Abolfazl  Ramezanpour &    Italy &  PhysRevE \\
        András  Lukács &  Hungary &  PhysRevE &        David D. Smith &             USA &  PhysRevA &              Alain  Barrat &   France &  PhysRevE \\
          Balázs  Rácz &  Hungary &  PhysRevE &     Douglas J. Ashton &  United Kingdom &  PhysRevE &     Alessandro  Vespignani &    Italy &  PhysRevE \\
         Brad R. Trees &      USA &  PhysRevB &        J. P. Saramäki &         Finland &  PhysRevE &            Alexei  Vázquez &      USA &  PhysRevE \\
       David G. Stroud &      USA &  PhysRevB &   Jukka Pekka  Onnela &  United Kingdom &  PhysRevE &        Andrea  Baronchelli &    Italy &  PhysRevE \\
            E.  Almaas &      USA &  PhysRevE &      Jussi M. Kumpula &         Finland &  PhysRevE &      Andrea  Lancichinetti &    Italy &  PhysRevE \\
      Erzsébet  Ravasz &  Romania &  PhysRevE &        János  Kertész &         Hungary &  PhysRevE &           Bruno  Gonçalves &   Brazil &  PhysRevD \\
        Gergely  Palla &  Hungary &  PhysRevE &             K.  Tucci &       Venezuela &  PhysRevE &                C. L. Zhang &      USA &  PhysRevB \\
          I.  Szakadát &  Hungary &  PhysRevE &        Kimmo K. Kaski &         Finland &  PhysRevB &        Claudio  Castellano &    Italy &  PhysRevE \\
       Illés J. Farkas &  Hungary &  PhysRevE &     Konstantin  Klemm &         Germany &  PhysRevE &            Daniele  Vilone &  Germany &  PhysRevE \\
         Imre  Derényi &  Hungary &  PhysRevE &          L.  Kullmann &         Hungary &  PhysRevE &           Dmitri  Krioukov &      USA &  PhysRevE \\
 M. Argollo De Menezes &      USA &      Other &      Mario G. Cosenza &       Venezuela &  PhysRevE &           Eric D. Kolaczyk &      USA &  PhysRevE \\
        R. V. Kulkarni &      USA &  PhysRevB &         Mark  Fricker &  United Kingdom &  PhysRevE &              Fabien  Viger &   France &  PhysRevE \\
         Tamás  Vicsek &  Hungary &  PhysRevE &  Martín G. Zimmermann &           Spain &  PhysRevE &          Filippo  Radicchi &    Italy &  PhysRevE \\
        V.  Saranathan &      USA &  PhysRevE &      Maxi  San Miguel &           Spain &  PhysRevA &        Francesca  Colaiori &    Italy &  PhysRevE \\
         Zoltán  Dezsö &      USA &  PhysRevE &         Mikko  Kivelä &         Finland &  PhysRevE &                Gary G. Yen &      USA &  PhysRevE \\
                       &          &           &       Neil F. Johnson &  United Kingdom &  PhysRevB &            José J. Ramasco &    Spain &  PhysRevE \\
                       &          &           &          R.  Toivonen &         Finland &  PhysRevE &            Luca  Dall’Asta &   France &  PhysRevE \\
                       &          &           &     Renaud  Lambiotte &         Belgium &  PhysRevE &         Ma Ángeles Serrano &    Spain &  PhysRevE \\
                       &          &           &           T. S. Evans &  United Kingdom &  PhysRevD &             Marián  Boguñá &    Spain &  PhysRevE \\
                       &          &           &    Timothy C. Jarrett &  United Kingdom &  PhysRevE &        Michel L. Goldstein &      USA &  PhysRevE \\
                       &          &           &     Víctor M. Eguíluz &           Spain &  PhysRevE &         Michele  Catanzaro &    Spain &  PhysRevE \\
                       &          &           &          X.  Castelló &           Spain &  PhysRevE &            Miguel A. Muñoz &    Spain &  PhysRevE \\
                       &          &           &                       &                 &           &              Nicola  Perra &    Italy &  PhysRevE \\
                       &          &           &                       &                 &           &        Philippe  Blanchard &  Germany &  PhysRevE \\
                       &          &           &                       &                 &           &  Romualdo  Pastor-Satorras &    Spain &  PhysRevE \\
                       &          &           &                       &                 &           &       S. Mehdi Vaez Allaei &     Iran &  PhysRevE \\
                       &          &           &                       &                 &           &           Santo  Fortunato &    Italy &  PhysRevE \\
                       &          &           &                       &                 &           &           Steven A. Morris &      USA &  PhysRevE \\
                       &          &           &                       &                 &           &               Tyll  Krüger &  Germany &  PhysRevE \\
                       &          &           &                       &                 &           &          Vittoria  Colizza &    Italy &  PhysRevE \\
                       &          &           &                       &                 &           &           Vittorio  Loreto &    Italy &  PhysRevE \\
\bottomrule
\end{tabular}
\caption{Authors in the three selected initial communities of the 2000s.}
\label{tab:aps_init_authors}
\end{sidewaystable}

\clearpage

\end{document}